# ALMA-IMF VIII – Combination of Interferometric Continuum Images with Single-Dish Surveys and Structural Analysis of Six Protoclusters

Daniel J. Díaz-González,[1] Roberto Galván-Madrid,[1] Adam Ginsburg,[2] Frédérique Motte,[3] Pierre Dell'Ova,[4,5] Stan Kurtz,[1] Nichol Cunningham,[3] Amelia M. Stutz,[6] Fabien Louvet,[3] Timea Csengeri,[7] Manuel Fernández-López,[8] Patricio Sanhueza,[9,10] Thomas Nony,[1] Rudy Rivera-Soto,[1] Rodrigo H. Álvarez-Gutiérrez,[6] Melanie Armante,[4,5] Melisse Bonfand,[11,7] Sylvain Bontemps,[7] Antoine Gusdorf,[4,5] and Hong-Li Liu[12]

[1]Instituto de Radioastronomía y Astrofísica, Universidad Nacional Autónoma de México, Morelia, Michoacán 58089, México.

[2]Department of Astronomy, University of Florida, PO Box 112055, USA.

[3]Univ. Grenoble Alpes, CNRS, IPAG, 38000 Grenoble, France.

[4]Laboratoire de Physique de l'École Normale Supérieure, ENS, Université PSL, CNRS, Sorbonne Université, Université de Paris, 75005, Paris, France.

[5]Observatoire de Paris, PSL University, Sorbonne Université, LERMA, 75014, Paris, France.

[6]Departamento de Astronomía, Universidad de Concepción, Casilla 160-C, 4030000 Concepción, Chile.

[7]Laboratoire d'astrophysique de Bordeaux, Univ. Bordeaux, CNRS, B18N, allée Geoffroy Saint-Hilaire, 33615 Pessac, France.

[8]Instituto Argentino de Radioastronomía (CCT-La Plata, CONICET; CICPBA), C.C. No. 5, 1894, Villa Elisa, Buenos Aires, Argentina

[9]National Astronomical Observatory of Japan, National Institutes of Natural Sciences, 2-21-1 Osawa, Mitaka, Tokyo 181-8588, Japan.

[10]Department of Astronomical Science, SOKENDAI (The Graduate University for Advanced Studies), 2-21-1 Osawa, Mitaka, Tokyo 181-8588, Japan.

[11]Departments of Astronomy and Chemistry, University of Virginia, Charlottesville, VA 22904, USA

[12]School of physics and astronomy, Yunnan University, Kunming, 650091, P. R. China.

## ABSTRACT

We present the combination of ALMA-IMF and single-dish continuum images from the Mustang-2 Galactic Plane Survey (MGPS90) at 3 millimeters and the Bolocam Galactic Plane Survey (BGPS) at 1 millimeter. Six and ten out of the fifteen ALMA-IMF fields are combined with MGPS90 and BGPS, respectively. The combination is made via the feathering technique. We used the dendrogram algorithm throughout the combined images, and performed further analysis in the six fields with combination in both bands (G012.80, W43-MM1, W43-MM2, W43-MM3, W51-E, W51-IRS2). In these fields, we calculated spectral index maps and used them to separate regions dominated by dust or free-free emission, and then performed further structural analysis. We report the basic physical parameters of the dust-dominated (column densities, masses) and ionized (emission measures, hydrogen ionization photon rates) structures. We also searched for multi-scale relations in the dust-dominated structures across the analyzed fields, finding that the fraction of mass in dendrogram leaves (which we label as "Leaf Mass Efficiency", LME) as a function of molecular gas column density follows a similar trend: a rapid, exponential-like growth, with maximum values approaching 100% in most cases. The observed behaviour of the LME with gas column is tentatively interpreted as an indicator of large star formation activity within the ALMA-IMF protoclusters. W51-E and G012.80 stand out as cases with comparatively large and reduced potential for further star formation, respectively.

## 1. INTRODUCTION

Massive star formation is a complex and highly dynamic process that plays a crucial role in the evolution of galaxies. Therefore, it is essential to understand how high-mass stars form and evolve (for a review, see Motte et al. 2018a).

The study of massive star and cluster formation has advanced considerably in recent years, thanks in part to Galactic surveys from infrared to radio wavelengths (e.g., Schuller et al. 2009; Molinari et al. 2010; Aguirre et al. 2011; Beuther et al. 2016; Brunthaler et al. 2021). Interferometric arrays such as the Atacama Large Millimeter/submillimeter Array (ALMA) provide high an-gular resolution, enabling detailed investigation of small-scale structures in molecular clouds located at distances of several kiloparsecs (e.g., Liu et al. 2015; Motte et al. 2018b; Sanhueza et al. 2019). However, interferometers are inherently insensitive to the large-scale emission, which is essential for understanding the overall structure of star-forming regions. Conversely, single-dish telescopes, such as those used in the Bolocam Galactic Plane Survey (BGPS, Aguirre et al. 2011; Ginsburg et al. 2013), and the MUSTANG Galactic Plane Survey (MGPS90, Ginsburg et al. 2020), observe the large-scale structures, but at the cost of lower angular resolution.

The majority of star formation occurs within stellar associations and clusters (Lada & Lada 2003), which appear to be largely substructured and hierarchical (e.g., Kumar et al. 2004; Gouliermis 2018). Individual (proto)stars and multiple stellar systems predominantly emerge from the densest peaks within the molecular-cloud hierarchy (e.g., Elmegreen 2008; Vázquez-Semadeni et al. 2019). Empirically, the relationships between less dense structures (full clouds of tens of pc in size to pc-scale clumps) and denser structures (cores and their aggregates, scales < 0.1 pc) is known to be complex and mutually influential, but their precise nature remains to be fully understood. To address this, it is important to investigate the hierarchical structures within star forming clouds. Recent studies have provided insights about cloud fragmentation (e.g., Palau et al. 2014; Sanhueza et al. 2019; Thomasson et al. 2022; Liu et al. 2023; Morii et al. 2023). One key aspect of fragmentation studies is the relation of the core mass function (CMF) to the origin of the stellar initial mass function (e.g., Könyves et al. 2015; Motte et al. 2018b; Suárez et al. 2021; Pouteau et al. 2022; Nony et al. 2023; Louvet et al. submitted). Other important aspects are the efficiency of the conversion of gas into stars (the star formation efficiency, SFE), and its predecessor, the efficiency at which gas moves from more diffuse to denser structures. A core formation efficiency (CFE) has been defined in previous literature to explore the latter (e.g., Louvet et al. 2014; Csengeri et al. 2017).

As molecular clouds evolve in time, feedback from star formation becomes increasingly significant. One of the most important sources of feedback is photoionization from massive stars reaching the main sequence, observed as pockets of free-free emission known as compact and ultracompact (UC) H ii regions (for reviews, see Churchwell 2002; Hoare et al. 2007). Within the ALMA-IMF sample, they become particularly relevant for the intermediate-stage and evolved protoclusters (Motte et al. 2022). Importantly, the associated free-free continuum emission can work as substantial "contam-inant" when interpreting millimeter continuum maps that focus exclusively on dust emission. Hence, the disentanglement of dust and free-free contributions in the millimeter photometry of massive star forming clouds is necessary to ensure a correct interpretation of observations. Therefore, in addition to studies of (dust-traced) cloud structure and fragmentation, it is important to quantify the presence of ionizing feedback from massive stars and their resulting H ii regions (e.g., Kurtz et al. 1994; Purcell et al. 2013; Ginsburg et al. 2016). These feedback processes are also a key factor in the self-regulation of star formation, with implications to the overall structure and evolution of the interstellar medium (e.g., Peters et al. 2010; Dale et al. 2014).

In this paper, we provide a feathering combination of continuum images at 3 mm and 1.3 mm from the ALMA-IMF Large Program with publicly available single-dish surveys, and perform a multi-scale analysis of the dust and free-free emission in the combined maps. In Section 2 we provide a brief description of the ALMA-IMF, BGPS, and MGPS90 data sets. In Section 3 we describe in detail the methodology employed for data combination. In Section 4 we proceed with the identification of hierarchical structure in the maps using the dendrogram algorithm, and present a detailed analysis of the six protoclusters for which we have data combination in both bands. This analysis involves calculating spectral index maps and utilizing them to characterize dust structures and H ii regions. In Section 5 we engage in a discussion regarding the efficiency of gas fragmentation between scales identified through the dendrogram analysis, and the possible relation between this and protocluster evolution. In Section 6 we provide our conclusions. The combined images and other data products are described in Appendix E, and distributed through the ALMA-IMF website[1] and a Zenodo[2] repository.

## 2. DATA

The data used in this paper come from three different projects and instruments: the ALMA-IMF Large Program (Motte et al. 2022), which used the Atacama Large Millimeter/submillimeter Array (ALMA); the Bolocam Galactic Plane Survey (Bolocam GPS Team 2020; Ginsburg et al. 2013), which used the Caltech Submillimeter Observatory (CSO); and the MUSTANG-2 Galactic Plane Survey (Ginsburg 2020; Ginsburg et al. 2020) at the Green Bank Telescope (GBT).

### 2.1. ALMA-IMF

---

The ALMA-IMF Large Program is a survey of fifteen massive protoclusters spanning a range of evolutionary stages, from very early star formation to having significant (proto)stellar feedback. The survey description and initial results can be found in Motte et al. (2022). The data reduction pipeline and interferometric continuum maps are presented in Ginsburg et al. (2022). Further pipeline development and the spectral line release are presented in Cunningham et al. (2023). ALMA-IMF has observations at 1.3 mm (Band 6, 216.2 GHz) and 3 mm (Band 3, 93.2 GHz), and its targets are at distances from 2 to 5.5 kpc. The physical resolutions ($\sim 2$ kau) and sensitivities are approximately homogeneous across the sample (Motte et al. 2022). Each ALMA-IMF mosaic has a slightly different central frequency shift (see details in the continuum release paper by Ginsburg et al. 2022).

ALMA-IMF released two sets of continuum images, labeled *cleanest* and *bsens* (Ginsburg et al. 2022). In the *cleanest* images most of the molecular line contamination was removed. The *bsens* images used the entire bandwidth from the observed spectral windows regardless of molecular line contamination. A modified *bsens-nobright* set of images that excludes the CO and a few other bright lines was also released. In this work we use these *bsens-nobright* images because they are the most analogous to the images obtained from single-dish bolometer cameras, in the sense that they use the entire available bandwidth but exclude bright line contamination. The ALMA-IMF images that we use achieve a $1\sigma$ noise level in the range of 0.05-0.7 mJy beam$^{-1}$ (see Table 1). We note that the released ALMA-IMF continuum images make use of only the main 12-m array. As mentioned in Section 3.4 of Ginsburg et al. (2022), it was found that images that included both the 7-m and 12-m arrays were noisier and had a degraded beam size and shape compared to those images using only the main ALMA array. The largest angular scale (LAS) recovered in the ALMA-IMF images is in the range of $10'' - 14''$ (Ginsburg et al. 2022).

The ALMA-IMF 12-m continuum data are available for download from the public Zenodo[3] repository. We used the primary-beam corrected images (`image.tt0.pbcor` file extension) because it is important that the interferometer images have the correct flux levels prior to combination.

The ALMA-IMF Band 6 images were combined with BGPS and the Band 3 images were combined with MGPS90. Five out of the fifteen ALMA-IMF fields do not have images in either MGPS90 or BGPS, and we do not consider them further in this paper. These regions are: G327.29, G328.25, G333.60, G337.92, G338.93. Table 1 summarizes the image combinations that we performed, listing the interferometric and single-dish beamsizes.

## 2.2. *BGPS*

BGPS (Aguirre et al. 2011) is a 1.1 mm continuum survey of 170 deg$^2$ of the Galactic plane visible from the northern hemisphere . The survey is contiguous over the range $-10.5° \leq l \leq 90.5°$, $|b| \leq 0.5°$. It achieves a non-uniform $1\sigma$ noise level in the range $11 - 53$ mJy beam$^{-1}$ . More details of the BGPS data release that we used can be found in Ginsburg et al. (2013).

This survey used Bolocam, which is the facility 144 bolometer array camera mounted at the Cassegrain focus of the 10.4 m mirror of the CSO on the summit of Mauna Kea. It used the filter configuration with a central frequency of 268 GHz (1.1 mm) and bandwidth of 46 GHz (Glenn et al. 2003). The bandpass is designed to reject emission from the CO($2 \to 1$) transition. We downloaded the data from the BGPS archive at https://irsa.ipac.caltech.edu (Bolocam GPS Team 2020). The archive contains: (1) the intensity map, and (2) a model of the signal at each position or alternatively the residual of subtracting the model from the survey map on the sky. The Bolocam field of view (FOV) is 7.5'. The pixel scale of the BGPS images is set to 7.2'', or 4.58 pixels across the 33'' half-power beamwidth (HPBW).

Like inteferometers, ground-based bolometer array observations filter out the largest angular scales. In particular, these observations are generally limited by the instantaneous FOV of the camera and the method of sky subtraction. The LAS recovered in the BGPS images is $\sim 2'$, which is a substantial improvement over the LAS provided by ALMA (see Section 2.1).

BGPS has 10 regions in common with ALMA-IMF: G008.67, G010.62, G012.80, W43-MM1, W43-MM2, W43-MM3, W51-E, W51-IRS2, G351.77, G353.41.

## 2.3. *MGPS90*

MGPS90 is an ongoing 3 mm continuum survey of the Galactic plane. Its pilot program covered about 7.5 deg$^2$ of the most prominent millimeter-bright regions between $0° < l < 50°$, $|b| < 0.5°$. The pilot survey achieved a typical $1\sigma$ depth of $1 - 2$ mJy beam$^{-1}$ . The MGPS90 pilot survey is described in greater detail in Ginsburg et al. (2020).

The survey uses MUSTANG-2 (Dicker et al. 2014), which is a 215-element bolometer array operating on the

| Region | ALMA B6 216.2 GHz [″ × ″, °] | ALMA B3 93.2 GHz [″ × ″, °] | BGPS 271.1 GHz [″] | MGPS90 91.5 GHz [″] |
|---|---|---|---|---|
| G008.67 | 0.72×0.59, -84 | – | 33 | – |
| G010.62 | 0.53×0.41, -78 | – | 33 | – |
| G012.80 | 1.10×0.71, 75 | 1.48×1.26, 88 | 33 | 9 |
| W43-MM1 | 0.51×0.36, -77 | 0.56×0.33, -74 | 33 | 9 |
| W43-MM2 | 0.54×0.42, -75 | 0.30×0.24, -73 | 33 | 9 |
| W43-MM3 | 0.53×0.45, 89 | 0.41×0.29, -85 | 33 | 9 |
| W51-E | 0.35×0.27, 26 | 0.29×0.27, 70 | 33 | 9 |
| W51-IRS2 | 0.51×0.44, -26 | 0.28×0.27, -60 | 33 | 9 |
| G351.77 | 0.90×0.67, 87 | – | 33 | – |
| G353.41 | 0.95×0.67, 85 | – | 33 | – |

**Table 1.** Columns are in order: field name, ALMA-IMF Band 6 synthesized HPBW and PA, ALMA-IMF Band 3 synthesized HPBW and PA, BGPS HPBW, and MGPS90 HPBW.

100 m Robert C. Byrd Green Bank Telescope (GBT). The central frequency is 90 GHz, and the bandwidth is 30 GHz (75-105 GHz). The MUSTANG-2 FOV is $4.25'$ and the beam HPBW is $9''$. The LAS recovered by MUSTANG-2 is $\sim 4.25'$. To recover structures at even larger angular scales, the pilot survey combined the MUSTANG-2 data with Planck images in the 100 GHz band (e.g., Csengeri et al. 2016; Lin et al. 2016; Abreu-Vicente et al. 2017). The Planck images at a central frequency of 104.225 GHz were scaled to a central frequency of 90.19 GHz assuming a spectral index $\alpha = 3$ (Ginsburg et al. 2020).

The MGPS90 pilot has 6 regions in common with ALMA-IMF: G012.80, W43-MM1, W43-MM2, W43-MM3, W51-E, W51-IRS2. We downloaded the data from their Dataverse[4] repository (Ginsburg 2020). The repository contains the MUSTANG-2 only maps and the MUSTANG-2 + Planck combined maps, the latter of which we use in this paper.

## 3. DATA COMBINATION

The benefits of combining interferometric and single-dish data have been widely recognized (e.g., Friesen et al. 2009; Koda et al. 2011; Liu et al. 2012; Galván-Madrid et al. 2013; Storm et al. 2014), as it allows to obtain images that possess the high angular resolution of interferometric data, while preserving the sensitivity to larger-scale emission from single-dish observations. Several techniques have been proposed to achieve this, including linear combination, feathering, and joint deconvolution (for reviews, see Stanimirovic 2002; Cotton 2017; Plunkett et al. 2023). Linear combination of interferometric and single-dish images is often performed directly in the image plane. Joint deconvolution, on the other hand, is a more sophisticated approach that seeks to find a common model of the sky by simultaneously deconvolving (e.g., CLEANing) both data sets, but it can be computationally intensive and complex to implement (Koda et al. 2019; Rau et al. 2019).

Feathering is an alternative Fourier-space technique that strikes a balance in complexity (e.g., Cotton 2017; Abreu-Vicente et al. 2017). It involves Fourier transforming both sets of images, combining them in Fourier space after appropriate flux scalings, and then transforming back to the image space. The convenience of feathering lies in its conceptual simplicity and computational efficiency compared to joint deconvolution, while still providing an accurate representation of the combined data. In this paper, we use the feathering technique as implemented in the `uvcombine`[5] Python package (Koch & Ginsburg 2022).

In this section we explain the combination of the interferometric and single-dish data. The procedure is schematized in Figure 1.

### 3.1. Preparation of FITS headers

Prior to any image manipulation, we homogenize the input FITS image headers and generate new FITS files without modifying their data. The original BGPS header contains the information about the rest frequency in the WAVELENG keyword (1.12 mm). In agreement with the ALMA standards, we generated a corresponding RESTFRQ keyword, and set it to 267.6718375 GHz. Keywords that describe the beam were missing in the MGPS90 headers. Therefore, we generated the keywords BMIN ($9''$), BMAJ ($9''$), and BPA ($0°$). Finally, the ALMA-IMF FITS files have the standard 4 dimensions (RA–SIN, DEC–SIN, FREQ, STOKES). Since we do not need the spectral and polarization information, we simplified these images by removing these two dummy axes. Furthermore, we change the RESTFRQ value to account for the fact that the exact central frequency of the ALMA images change by a few MHz from field to field. To determine RESTFRQ, we take into account that the intrinsic spectral index of the emitting source is not flat. We use the values in Table D.1 of Ginsburg et al. (2022) for a spectral index $\alpha$ = 3.5, that corresponds to optically thin dust emission with an emissivity idex of $\beta$ =1.5, well suited for star forming regions (Andre et al. 1993; Juvela et al. 2015).

### 3.2. Cropping, regridding and reprojection

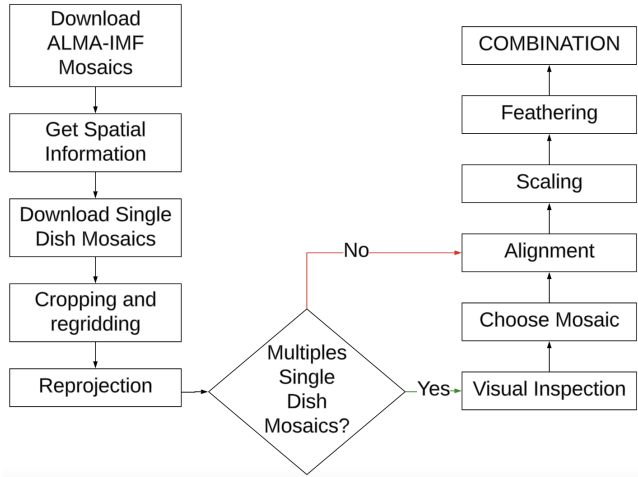

**Figure 1.** Flowchart of the procedure to combine the interferometric and single-dish data.

To combine images from different instruments, we need to homogenize their geometric properties and spatial information. We begin by obtaining the single-dish images centered on the ALMA-IMF fields, and fixing the FITS headers as described in Section 3.1. We then process the single-dish files with the `reproject_interp` task from the `reproject` Python package (Robitaille et al. 2020). This is performed in order to extract the interferometric FOVs (cropping), to get the interferometric matrix shape and pixel size (regridding) and to reproject to the interferometric coordinate system for every single-dish image and its matching interferometric mosaic. We illustrate the procedure in Figure 2.

It should be noted that in the BGPS archive, there can occasionally exist multiple FITS files corresponding to a given ALMA-IMF Band 6 field. In these cases we perform an additional visual inspection of the extracted FOVs and choose those data where the ALMA-IMF field is better centered and has lower noise.

### 3.3. Alignment

Once the single-dish images have been processed to extract the ALMA-IMF FOV's, we compare the single-dish versus interferometric astrometry. The astrometric accuracy of the images is a fraction of the respective beamsize (see Table 1), thus the interferometric astrometric accuracy is expected to be better. Therefore, we use the interferometric mosaic to set the alignment. The `image_registration`[6] Python package has been used to align every single-dish image to the matching interferometric mosaic. We use the `chi2_shift` task to get the offsets. This task calculates the offsets between two images using the DFT[7] upsampling method. For this, we use the ALMA-IMF primary beam to mask the pixels with response $\leq 0.5$ to avoid artifacts from the mosaic edges, and then we convolve the ALMA-IMF image to the single-dish beam.

[6] https://image-registration.readthedocs.io/en/latest/

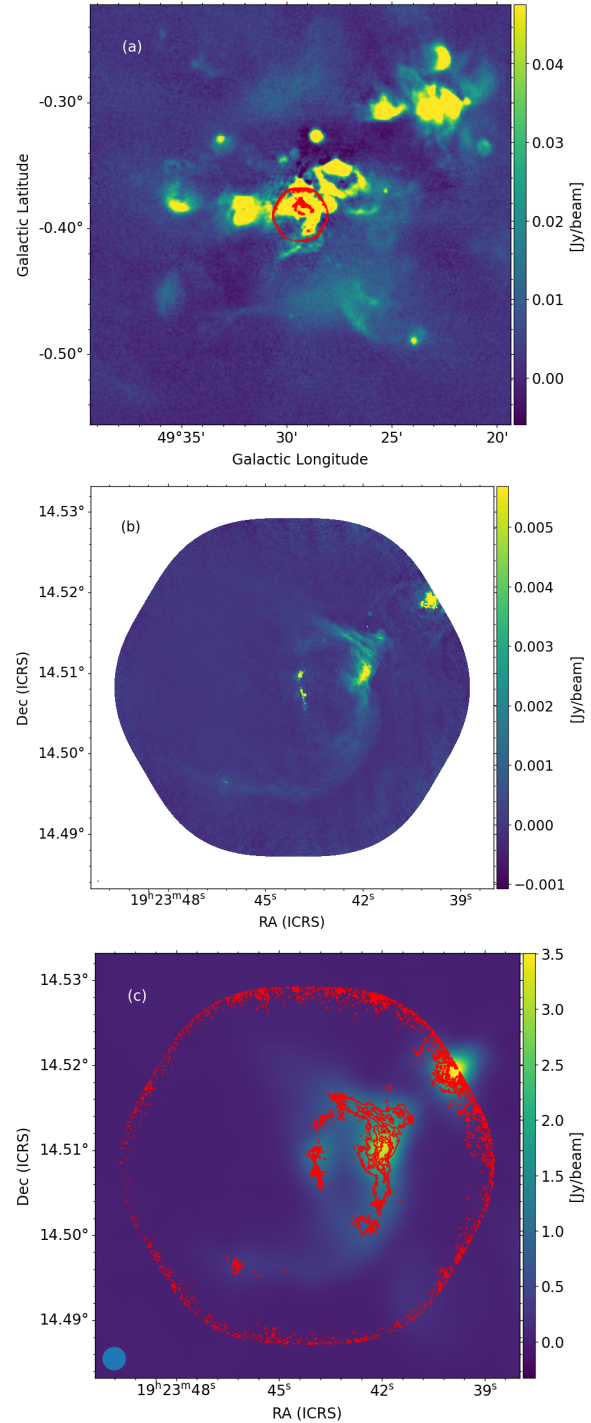

**Figure 2.** Reprojection procedure for W51-E. From top to bottom: *(a)* MGPS90 image with ALMA-IMF Band 3 mosaic outline in red contours before reprojection. *(b)* ALMA-IMF Band 3 mosaic. *(c)* ALMA-IMF Band 3 FOV in red contours and MGPS90 after reprojection in colors. Panels *(b)* and *(c)* show the ALMA-IMF Band 3 (barely noticeable red circle) and MGPS90 (larger blue circle) beam in their bottom-left corners.

We use the calculated offsets to shift the coordinates of the single-dish images and generate new FITS files with the corrected coordinates. We illustrate the results of the alignment procedure for the case of W51-E at 3 mm and W43-MM3 at 1 mm in Figures 3 and 4, respectively. These two examples are among the largest offsets found across the sample. The alignment corrections for all the combinations are reported in Table 2. In all cases they are a small fraction of the single-dish beamsize.

| Field | BGPS RA ['']| BGPS Dec ['']| MGPS90 RA ['']| MGPS90 Dec [''] |
|-------|--------------|--------------|----------------|------------------|
| G008.67 | 1.43±0.02 | 5.35±0.02 | - | - |
| G010.62 | -2.68±0.01 | -1.56±0.01 | - | - |
| G012.80 | -0.92±0.01 | 2.99±0.01 | 1.48±0.05 | 0.36±0.04 |
| W43-MM1 | -3.01±0.01 | 2.19±0.01 | -0.98±0.31 | -0.70±0.29 |
| W43-MM2 | -3.86±0.02 | 1.11±0.02 | -0.00±0.01 | -0.00±0.01 |
| W43-MM3 | -3.60±0.04 | 0.31±0.04 | -0.61±0.11 | -0.28±0.11 |
| W51-E | -3.53±0.00 | 1.60±0.01 | -6.49±0.01 | -0.55±0.01 |
| W51-IRS2 | 0.01±0.00 | -0.01±0.01 | -6.06±0.01 | -0.11±0.01 |
| G351.77 | 3.33±0.01 | 4.22±0.01 | - | - |
| G353.41 | 4.61±0.03 | 1.46±0.02 | - | - |

**Table 2.** Astrometric corrections for each region. We show the shifts of the single-dish images required for alignment with respect to the ALMA-IMF mosaics. Columns 2 and 3 show shifts and their errors for the BGPS images. Columns 4 and 5 show the shifts and their errors for MGPS90.

### 3.4. Feathering

Before combining the single-dish and interferometric data, we check for flux calibration differences between them. Interferometers and single-dish are not sensitive to all angular scales on the sky, both are limited in their measurable smallest (SAS) and largest (LAS) angular scales. The SAS of a single-dish telescope is determined by its aperture diameter, while the SAS of interferometers is determined primary by their longest baselines. In contrast, the LAS of an interferometer will generally be smaller than the LAS of the single-dish, given the lack of zero-spacing information. The LAS of an interferometer is given by its shortest baselines, but its value is not as sharply defined as is usually considered, and rather linked to the density of the Fourier sampling of these shortest baselines (see, e.g., Ginsburg et al. 2022).

The observable effect is that interferometers recover significantly less flux than single-dishes for the extended emission approaching or exceeding their LAS. As mentioned before, the LAS of single-dish observations is not infinite either, but for bolometer observations it will depend on the details of the sky subtraction, often linked to the instantaneous field-of-view (FOV) of the bolometer array (see, e.g., Ginsburg et al. 2013; Csengeri et al. 2016; Lin et al. 2016; Abreu-Vicente et al. 2017).

However, the flux measured by the interferometer and the single-dish in the angular scales that they have in common should be the same. We use this to calibrate the flux scale of the single-dish data, applying a correction factor $f_{amp}$. We determine the interferometric LAS from the typical 95$^{th}$ percentile of the baseline histograms presented in Ginsburg et al. (2022). From this, the LAS of the ALMA-IMF-B3 mosaics is $\approx 14''$, and for the ALMA-IMF-B6 mosaics it is $\approx 10''$. To define the single-dish SAS, for MGPS90 we take $10''$, which is slightly larger than the nominal beam size of $9''$ quoted in Ginsburg et al. (2020). This allows us to take into account potential variations in the actual beamsize among different observations. Therefore, the range of overlapping scales to determine $f_{amp}$ for the combination at 3 mm is approximately from $10''$ to $14''$. In contrast, BGPS and the ALMA-IMF B6 mosaics only have the residual overlap of the small percentage of baselines that sample scales $\gtrsim 10''$. For this reason, the BGPS data have not been scaled.

Furthermore, the single-dish and interferometric bandwidths are centered at different frequencies, and each ALMA-IMF mosaic has a slightly different frequency shift (see Section 2.1). To scale the flux between different frequencies, we use the relation $S_\nu \propto \nu^{3.5}$. The frequency scaling factor $f_\nu$ is slightly different for each combination of field and band (see Table 3).

We use the feathering algorithm, as implemented in the `uvcombine` package (Koch & Ginsburg 2022), to combine the single-dish and interferometric data. Our pipeline is as follows:

1. The ALMA-IMF images are masked out where the primary beam response is < 0.3.

2. The frequency scaling factor $f_\nu$ is calculated and applied to the single-dish images.

3. The flux units of both single-dish and interferometric images are converted from Jy/beam to Jy/px.

4. A Fast Fourier Transform (FFT) is applied to both images.

---

[7] Manuel Guizar (2022). Efficient subpixel image registration by cross-correlation, MATLAB Central File Exchange.

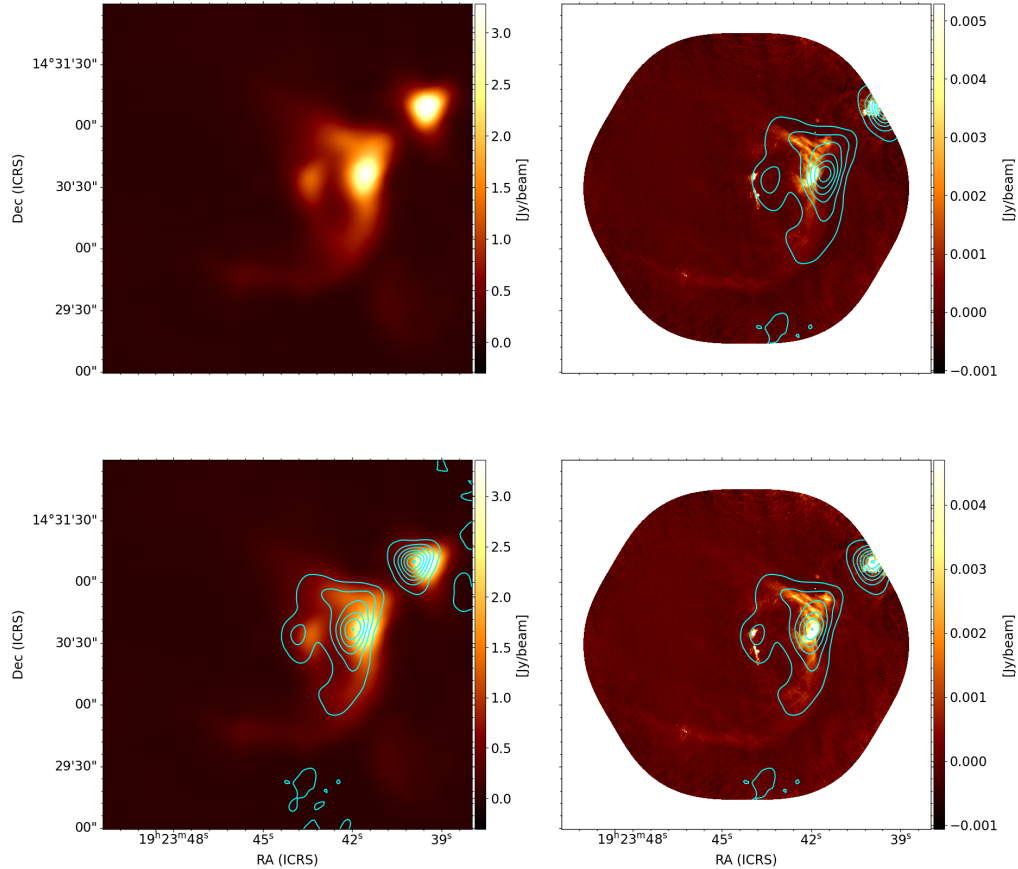

**Figure 3.** Alignment correction ($\Delta$RA $= -6.49''$, $\Delta$Dec $= -0.55''$) for the MGPS90 and ALMA-IMF B3 combination for W51-E. *Top-Left:* MGPS90 original (unshifted) data. *Top-Right:* MGPS90 data with original (unshifted) astrometry overlaid as contours on the ALMA-IMF image.. *Bottom-Left:* MGPS90 data with shifted astrometry overlaid as contours on the MGPS90 original image. *Bottom-Right:* MGPS90 data with shifted astrometry overlaid as contours on the ALMA-IMF image.

5. The average amplitude ratio for the common scales, $f_{\mathrm{amp}}$, is calculated for the ALMA-IMF B3 + MGPS90 combination (see Table 3). Due to the lack of overlapping scales in Band 6, for the combination of ALMA-IMF B6 + BGPS we set $f_{\mathrm{amp}}$ to 1. Figure 5 shows the calculation and comparison before and after the application of $f_{\mathrm{amp}}$ for the W51-E Band 3 combination.

6. The factor $f_{\mathrm{amp}}$ is applied to the single-dish FFT amplitudes.

7. We perform the data combination in the Fourier domain, following:

$$C(u,v) = w'\mathrm{FFT}[I_{\mathrm{int}}(x,y)] + w''\mathrm{FFT}[AI_{\mathrm{sd}}(x,y)],$$
(1)

where $C(u,v)$ is the combined data in Fourier space, $A = f_{\mathrm{amp}} \times f_\nu$, $I_{\mathrm{int}}(x,y)$ is the interferometric image, $I_{\mathrm{sd}}(x,y)$ is the single-dish image, $w''$ is the weighting for the low-resolution image, and $w' = 1 - w''$ is the weighting for the high-resolution image. Within the `uvcombine` package, we use the function `feather_kernel` to calculate $w'$ and $w''$, and `fftmerge` to calculate the combination.

8. The inverse FFT is applied to $C(u,v)$ to get the combined image $I_{\mathrm{comb}}(x,y)$ in Jy/px.

9. The flux units of the combined images are converted back from Jy/px to Jy per interferometric beam.

Figure 5 illustrates the flux scaling procedure. Table 3 shows the corresponding frequency, the single-dish flux prior to any scaling ($S_0$), the frequency scaling fac-tor ($f_\nu$), the frequency-scaled single-dish flux ($S_\nu$), the common-scales factor ($f_{\text{amp}}$), and the final scaled single-dish flux ($S_{\text{amp}}$) for each field. Fluxes in Table 3 correspond to all pixels over the footprint of the ALMA-IMF FOV. Table 4 shows relevant parameters for the Band 3 and Band 6 combined images. The fluxes reported in this table are only for pixels above a $5\sigma$ noise threshold. We note that the combination generally does not degrade the initial noise level of the ALMA images, with a slight improvement in some cases (see Table 4). The combination procedure preserves the initial angular resolution of ALMA images.

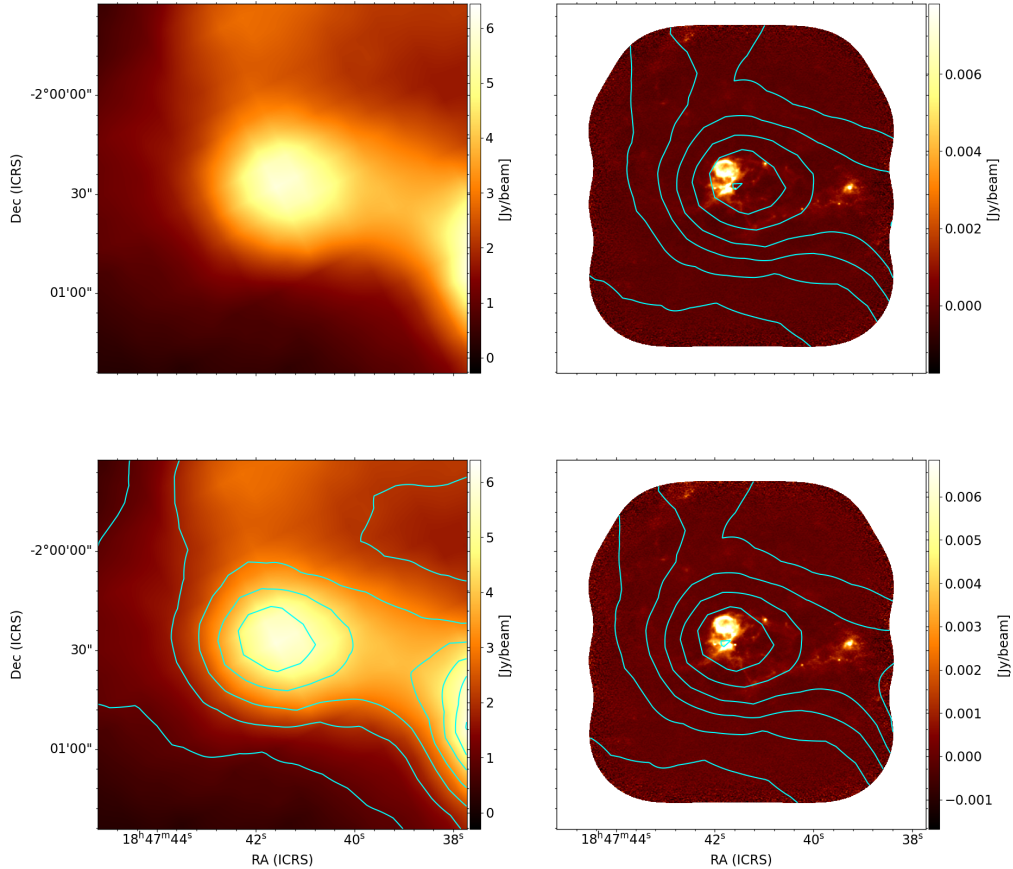

**Figure 4.** Alignment correction ($\Delta\text{RA} = -3.60''$, $\Delta\text{Dec} = 0.31''$) for the BGPS and ALMA-IMF B6 combination for W43-MM3. *Top-Left:* Original BGPS astrometry in colors. *Top-Right:* Original BGPS astrometry (contours) over ALMA-IMF-B6 astrometry (colors). *Bottom-Left:* Shifted BGPS astrometry (contours) over original BGPS astrometry (colors). *Bottom-Right:* Shifted BGPS astrometry (contours) over ALMA-IMF-B6 astrometry (colors).

Figures 11 and 12 in Appendix A show the resulting combined images for each band. As expected, the brightest emission remains without noticeable changes because it corresponds to structures that are compact enough to be well recovered by the interferometer. How-ever, the total recovered fluxes increase by factors in the range $\times 1.3$ to $\times 4$ compared to the interferometric images (see Table 4). This extra flux is faint because it is distributed over many resolution elements across the maps. The extended flux recovered after combination with intensity levels between $5\sigma$ and $10\sigma$, compared to the interferometric data, increased by factors from $\times 1.4$ to $\times 12.4$ for Band 3, and from $\times 1.4$ to $\times 6.4$ for Band 6.

## 4. ANALYSIS

Disentangling between free-free and dust emission is crucial to determine the mass content in structures within molecular clouds. In this section, we employ the image combinations to identify structures within each band, using the dendrogram algorithm (Rosolowsky et al. 2008). Furthermore, these image combinations are used to generate spectral index maps for the six regions with combinations in both bands. These spectral index

(a) Before frequency and flux scaling.

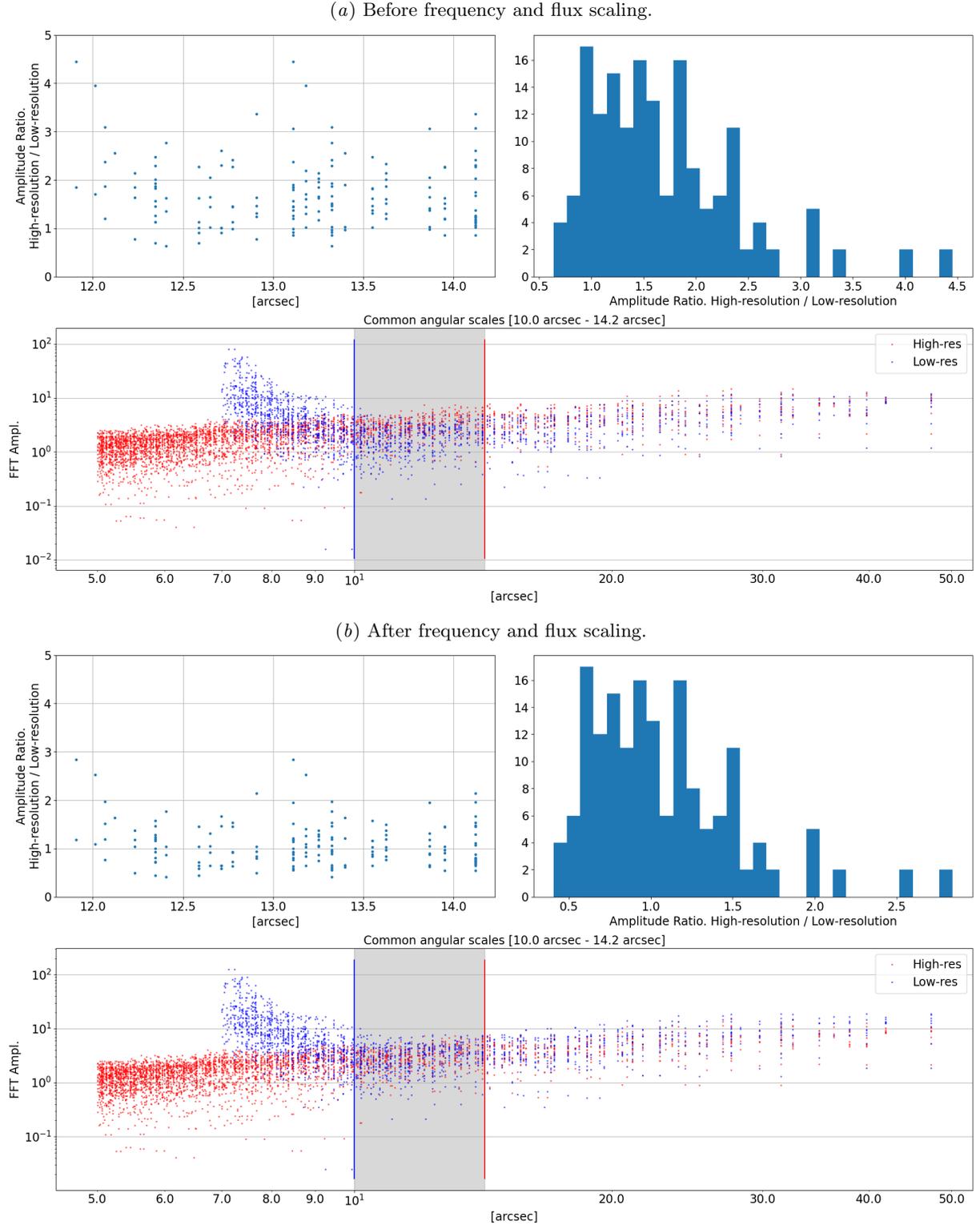

(b) After frequency and flux scaling.

**Figure 5.** Flux scaling procedure for W51-E in ALMA-IMF Band 3 and MGPS90. For each of the top (*a*, before scaling) and bottom (*b*, after scaling) images: *Top-left* – Ratio between FFT amplitudes for common angular scales between interferometric and single-dish data. *Top-right* – Histogram of these ratios. *Bottom* – FFT amplitudes from the interferometer (red dots) and single-dish (blue dots). The gray shaded area shows the common scales.

| Region | $\nu_{int}$ [GHz] | $\nu_{sd}$ [GHz] | $S_0$ [Jy] | $f_\nu$ | $S_\nu$ [Jy] | $f_{amp}$ | $S_{amp}$ [Jy] |
|---|---|---|---|---|---|---|---|
| MGPS90 + ALMA-IMF-B3 | | | | | | | |
| W43-MM1 | 100.686 | 90.19 | 5.76 | 1.47 | 8.47 | 0.60 | 5.10 |
| W43-MM2 | 100.682 | 90.19 | 6.72 | 1.47 | 9.87 | 0.77 | 7.60 |
| W43-MM3 | 100.682 | 90.19 | 4.15 | 1.47 | 6.09 | 0.73 | 4.47 |
| G012.80 | 100.710 | 90.19 | 33.41 | 1.42 | 47.37 | 0.74 | 34.94 |
| W51-E | 100.706 | 90.19 | 39.51 | 1.42 | 56.02 | 1.14 | 63.72 |
| W51-IRS2 | 100.704 | 90.19 | 37.90 | 1.42 | 53.74 | 0.97 | 52.17 |
| BGPS + ALMA-IMF-B6 | | | | | | | |
| G008.67 | 228.951 | 267.67 | 22.14 | 0.58 | 12.81 | 1.00 | 12.81 |
| G010.62 | 228.971 | 267.67 | 31.22 | 0.58 | 18.08 | 1.00 | 18.08 |
| G012.80 | 228.941 | 267.67 | 96.65 | 0.58 | 55.93 | 1.00 | 55.93 |
| G351.77 | 228.982 | 267.67 | 59.67 | 0.58 | 34.55 | 1.00 | 34.55 |
| G353.41 | 229.000 | 267.67 | 56.19 | 0.58 | 32.54 | 1.00 | 32.54 |
| W43-MM1 | 229.494 | 267.67 | 31.11 | 0.58 | 18.16 | 1.00 | 18.16 |
| W43-MM2 | 228.901 | 267.67 | 24.83 | 0.58 | 14.36 | 1.00 | 14.36 |
| W43-MM3 | 228.902 | 267.67 | 17.09 | 0.58 | 9.88 | 1.00 | 9.88 |
| W51-E | 228.938 | 267.67 | 119.14 | 0.58 | 68.92 | 1.00 | 68.92 |
| W51-IRS2 | 228.931 | 267.67 | 98.68 | 0.58 | 57.09 | 1.00 | 57.09 |

**Table 3.** $\nu_{int}$: central frequency of the interferometric ALMA-IMF image. $\nu_{sd}$ central frequency of the MGPS90/BGPS image. $S_0$: original flux from single-dish image. $f_\nu$: frequency scaling factor. $S_\nu$ single-dish flux after applying $f_\nu$. $f_{amp}$: common-scale scaling factor (fixed to 1 for BGPS, see Section 3.4). $S_{amp}$ single-dish flux after applying both $f_{amp}$ and $f_\nu$.

maps allow us to separate dust-dominated from free-free dominated areas, which in turn enable further analysis in Section 5. The structures detected in Band 6 and dominated by dust emission are used to measure the efficiency of gas conversion between intermediate and small scales. The structures detected in Band 3 and dominated by free-free emission are then used to measure the amount of photoionization and propose proxies for the evolution of star forming molecular clouds.

### 4.1. Structure identification

We use the `astrodendro` Python package to identify hierarchical structure in the combined images. This package implements the dendrogram algorithm as described in Rosolowsky et al. (2008). The algorithm traces hierarchical structure over different scales and labels them as *trunks* for the largest structures, *branches* for the intermediate ones, *leaves* for the smallest structures without further substructure. In this work we use the dendrogram hierarchy to explore multi-scale relations between extended and compact structures in some of the protocluster clumps of the ALMA-IMF sample.

*Image noise* – In order to run `astrodendro`, we need a measure of the noise in the image. In this work we use the median absolute deviation (MAD) of pixel intensities as a noise measurement, because it is robust to out-lier points. However, our combined images can be dominated by signal over a large fraction of the maps. Therefore, we define a recursive MAD$^{\leftarrow\rho}$ to reject extended emission in the maps and use a mask that contains almost only noise. The iterative procedure to calculate MAD$^{\leftarrow\rho}$ is as follows: the initial MAD is calculated for the full combined image, then a mask is applied that rejects pixels with values $> 2 \times$ MAD, then the MAD is recalculated. The stopping criterion is when the difference between successive iterations is less than 1%, which happens after a few iterations in all cases. The MAD$^{\leftarrow\rho}$ values are reported in Table 4.

*Generating the dendrograms* – The `astrodendro` package (Robitaille et al. 2019) has three free parameters. We associate each of them with one physical quantity:

- `min_npix` is the minimum number of pixels needed for a leaf to be considered an independent entity. We set it conservatively to $1.5 \times$ the beam area in pixels. A source that is truly point-like would formally have an observed size of one beam.

- `min_value` is the minimum intensity value considered when computing the dendrogram tree. Values below this threshold will be ignored. We set it to $5 \times$ MAD$^{\leftarrow\rho}$ in order to avoid spurious detections.

- `min_delta` is the minimum relative height a branch or leaf needs in order to be considered an independent entity. We set it to $3 \times$ MAD$^{\leftarrow\rho}$.

The values for `min_value` and `min_delta` were optimized from visual inspection of the outputs. We inspected a range of $\times[3, 5, 7, 9]$ in both parameters. The selected values are conservative in the sense that a relatively large ($5\sigma$ rather than $3\sigma$) `min_value` ensures that residual sidelobes in the maps are not considered as real sources. This extra rejection threshold is not needed for `min_delta`, which is concerned with structures of higher intensity in the hierarchy. Overall, our selected parameters should produce a catalogue of structures that contains few spurious sources, but that is also far from complete at faint fluxes. The dendrogram properties of the combined images are given in Table 5. We distinguish three types of structures: trunks, corresponding to structures that themselves have internal substructures; leaves, corresponding to the smallest substructures belonging to a trunk; and isolated structures, which are structures without internal substructures and do not belong to any trunk.

*Fluxes and background* – Once the dendrogram structures are identified, we compute the background-subtracted fluxes of the leaves in Jansky units. We es-timate the local background of each leaf by measuring the average intensity of each trunk excluding all of its leaves. This background intensity is then subtracted over the the mask of each leaf. Flux errors are propagated using the image noise determined as described above. All errors reported in this paper are propagated analytically.

| | | | | MGPS90 + ALMA-IMF-B3 | | | |
|---|---|---|---|---|---|---|---|
| Field | From | RESTFRQ [GHz] | Flux [Jy] | MAD$^{-\rho}$ [Jy beam$^{-1}$] | Max [Jy beam$^{-1}$] | Min [Jy beam$^{-1}$] | DR |
| W43-MM1 | MGPS90 | 90.190 | 3.45 | 0.00590 | 0.2546 | -0.0062 | 43.4 |
| | ALMA-IMF | 100.686 | 1.15 | 0.00006 | 0.0186 | -0.0011 | 293.6 |
| | COMB | 100.686 | 1.90 | 0.00006 | 0.0187 | -0.0011 | 330.4 |
| W43-MM2 | MGPS90 | 90.190 | 7.53 | 0.00120 | 0.7888 | -0.0022 | 658.8 |
| | ALMA-IMF | 100.682 | 2.93 | 0.00005 | 0.0050 | -0.0009 | 110.5 |
| | COMB | 100.682 | 3.85 | 0.00006 | 0.0050 | -0.0009 | 117.6 |
| W43-MM3 | MGPS90 | 90.190 | 4.35 | 0.00100 | 0.7509 | -0.0021 | 761.5 |
| | ALMA-IMF | 100.682 | 2.13 | 0.00005 | 0.0059 | -0.0012 | 108.6 |
| | COMB | 100.682 | 2.86 | 0.00005 | 0.0060 | -0.0011 | 116.9 |
| G012.80 | MGPS90 | 90.190 | 34.94 | 0.0010 | 7.1270 | -0.0008 | 7486.9 |
| | ALMA-IMF | 99.655 | 20.48 | 0.00047 | 0.8477 | -0.0076 | 1796.0 |
| | COMB | 99.655 | 32.98 | 0.00030 | 0.8492 | -0.0062 | 2791.9 |
| W51-E | MGPS90 | 90.190 | 63.69 | 0.0015 | 6.2763 | -0.0088 | 4181.4 |
| | ALMA-IMF | 99.651 | 20.15 | 0.00011 | 0.4058 | -0.0014 | 3803.1 |
| | COMB | 99.651 | 45.46 | 0.00006 | 0.4060 | -0.0012 | 6549.7 |
| W51-IRS2 | MGPS90 | 90.190 | 51.37 | 0.0069 | 5.3578 | -0.0289 | 775.2 |
| | ALMA-IMF | 99.650 | 20.70 | 0.00015 | 0.3492 | -0.0043 | 2350.8 |
| | COMB | 99.650 | 30.49 | 0.00011 | 0.3494 | -0.0042 | 3059.6 |
| | | | | BGPS + ALMA-IMF-B6 | | | |
| G008.67 | BGPS | 267.672 | 12.8106 | 0.0633 | 6.3282 | 0.2118 | 100.0 |
| | ALMA-IMF | 228.951 | 4.2395 | 0.0004 | 0.2258 | -0.0076 | 599.2 |
| | COMB | 228.951 | 10.0642 | 0.0003 | 0.2261 | -0.0073 | 667.0 |
| G010.62 | BGPS | 267.672 | 18.0760 | 0.1107 | 12.3397 | 0.3836 | 111.5 |
| | ALMA-IMF | 228.971 | 10.0090 | 0.0002 | 0.3990 | -0.0030 | 1979.7 |
| | COMB | 228.971 | 16.2151 | 0.0002 | 0.3991 | -0.0028 | 2402.4 |
| G012.80 | BGPS | 267.672 | 55.9251 | 0.1097 | 18.9550 | 0.3923 | 172.8 |
| | ALMA-IMF | 228.941 | 28.6975 | 0.0007 | 0.4176 | -0.0102 | 607.6 |
| | COMB | 228.941 | 51.8082 | 0.0004 | 0.4189 | -0.0090 | 986.5 |
| G351.77 | BGPS | 267.672 | 34.5482 | 0.0116 | 20.5441 | 0.1225 | 1768.0 |
| | ALMA-IMF | 228.982 | 12.2675 | 0.0006 | 0.6794 | -0.0113 | 1185.6 |
| | COMB | 228.982 | 28.9014 | 0.0004 | 0.6801 | -0.0106 | 1549.5 |
| G353.41 | BGPS | 267.672 | 32.5445 | 0.0267 | 10.3775 | 0.1460 | 388.4 |
| | ALMA-IMF | 229.000 | 7.4824 | 0.0013 | 0.1145 | -0.0278 | 88.4 |
| | COMB | 229.000 | 28.1703 | 0.0012 | 0.1155 | -0.0269 | 99.7 |
| W43-MM1 | BGPS | 267.672 | 18.1575 | 0.1533 | 9.7592 | 0.3719 | 63.7 |
| | ALMA-IMF | 229.494 | 7.4791 | 0.0002 | 0.3942 | -0.0029 | 2008.7 |
| | COMB | 229.494 | 14.9999 | 0.0002 | 0.3944 | -0.0028 | 2304.8 |
| W43-MM2 | BGPS | 267.672 | 14.3585 | 0.1578 | 5.5727 | 0.4296 | 35.3 |
| | ALMA-IMF | 228.901 | 3.6196 | 0.0002 | 0.1901 | -0.0038 | 841.6 |
| | COMB | 228.901 | 13.0478 | 0.0002 | 0.1904 | -0.0035 | 1235.7 |
| W43-MM3 | BGPS | 267.672 | 9.8850 | 0.0105 | 3.4572 | 0.3066 | 329.8 |
| | ALMA-IMF | 228.902 | 2.9712 | 0.0001 | 0.0573 | -0.0022 | 460.8 |
| | COMB | 228.902 | 9.9378 | 0.0001 | 0.0575 | -0.0020 | 700.9 |
| W51-E | BGPS | 267.672 | 68.9207 | 0.2704 | 31.7989 | 1.6854 | 117.6 |
| | ALMA-IMF | 228.919 | 39.6971 | 0.0005 | 0.4498 | -0.0077 | 921.9 |
| | COMB | 228.919 | 71.0830 | 0.0004 | 0.4502 | -0.0073 | 1122.7 |
| W51-IRS2 | BGPS | 267.672 | 57.0932 | 0.2459 | 28.3850 | 1.2307 | 115.4 |
| | ALMA-IMF | 228.931 | 29.6458 | 0.0002 | 0.9407 | -0.0030 | 3963.6 |
| | COMB | 228.931 | 63.0876 | 0.0002 | 0.9417 | -0.0020 | 5801.8 |

**Table 4.** Image parameters for the combinations of MGPS90 + Band 3 and BGPS + Band 6. From left to right: field name, central frequency of the input/output image, flux measured over the pixels above $5\sigma$, iteratively-calculated median absolute deviation ($\sigma$ noise, see Section 4.1), maximum and minimum intensity, dynamic range (positive maximum intensity divided by MAD$^{-\rho}$).

*Homogeneous analysis of the six fields in both bands* – We use the identified dendrogram tree together with the spectral index maps to distinguish between dust-dominated and free-free-dominated structures (see Section 4.2). To accomplish this, we must generate spectral index maps. Hence, we convolve each of the regions for which we have both 1 mm and 3 mm feathered images to their common circular beam. The dendrogram identification is run in the convolved images, and the results that we report for these six fields (G012.80, W43-MM1, W43-MM2, W43-MM3, W51-IRS2, and W51-E) are after convolution. The actual beams that were used for all fields are reported in Table 5.

Table 5 also presents the number of dendrogram structures identified for each field. Figure 6 presents the histograms of effective diameters for the dendrogram structures identified in all fields. Most of the leaves have effective diameters < 0.1 pc, and none of them is larger than 0.2 pc. Similarly, most of the trunks have sizes larger than ≳ 0.1 pc. In Band 6, the average diameter of trunks is 0.238 pc, the average diameter of leaves is 0.039 pc, and the average diameter of isolated elements is 0.041 pc. In Band 3, the average diameter of trunks is 0.416 pc, the average diameter of leaves is 0.042 pc, and the average diameter of isolated elements is also 0.042 pc. Table 6 presents the statistics for each region.

The main purpose of using dendrograms in this paper is to explore multi-scale relations in the hierarchy of star forming structures. The previous analysis shows that the dendrogram hierarchy naturally separates objects above and below the sizescale of "massive dense cores" (MDCs, 0.1 pc, see Table 1 of Motte et al. 2018a). The dendrogram trunks with sizes between 0.1 and 1 pc will then correspond to star forming "clumps", as long as their emission is dominated by dust (see Section 4.2). The dendrogram leaves, however, should not be interpreted as a catalogue of individual pre- and protostellar cores. One consequence of our definition of `min_npix` is that our smallest sources are similar in size to the largest cores extracted using the `getsf` algorithm (Men'shchikov 2021). We refer to Louvet et al. (submitted) for an analysis of the individual star-forming cores in ALMA-IMF fields.

### 4.2. *Separation of thermal dust and free-free emission*

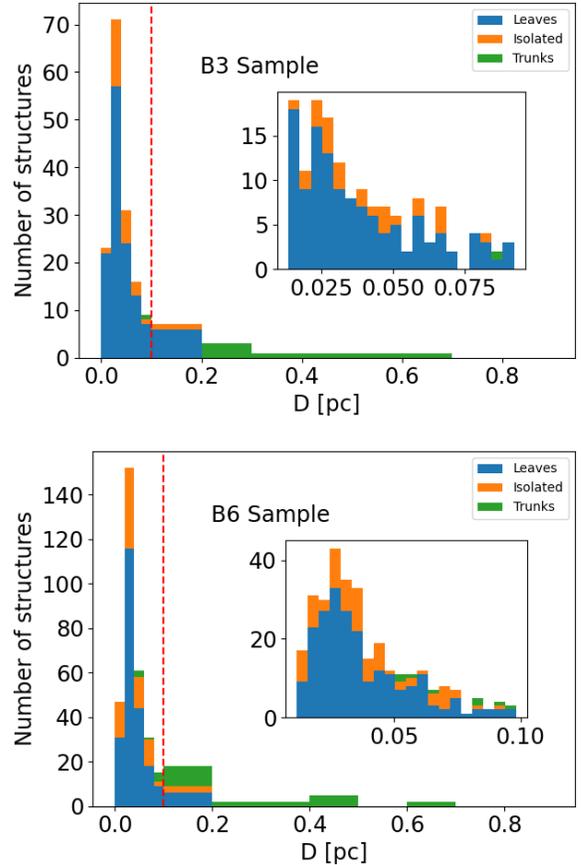

**Figure 6.** Histograms of effective diameters for the entire sample of dendrogram structures. The top and bottom panels show the structures in Band 3 and Band 6, respectively. The vertical red line marks the 0.1 pc threshold that approximately separates small and large scales. The total number of leaves (blue), structures without internal substructures (orange), and trunks (green) are shown, corresponding to the size indicated by the bin. These numbers refer to the total sample, summing across all fields. Insets highlight structures with diameters < 0.1 pc.

Different mechanisms are responsible for the emission in the ALMA-IMF maps, mainly free-free emission from H II regions and interstellar dust emission. The ALMA-IMF fields were selected in order to be a representative sample of massive protoclusters in different evolutionary stages (Motte et al. 2022), therefore the younger fields are entirely dominated by dust emission, whereas free-free emission increasingly contributes in the maps of the more evolved fields, mainly in Band 3.

To separate emission mechanisms, we use the spectral index maps in the convolved images of the six available fields. The spectral index $\alpha$ is defined as:

$$\alpha = \frac{\ln\left(I_{\mathrm{B6}}/I_{\mathrm{B3}}\right)}{\ln\left(\nu_{\mathrm{B6}}/\nu_{\mathrm{B3}}\right)}, \qquad (2)$$

| Field | Beam | MAD$^{-\rho}$ | $N_{\rm thr}$ | Total | Trunks | Leaves | Isolated |
|---|---|---|---|---|---|---|---|
| | [$'' \times ''$,°] | [Jy beam$^{-1}$] | | | | | |
| BPGS + ALMA-IMF-B6 | | | | | | | |
| G008.67 † | 0.72×0.59, -81 | 3.39e-04 | - | 21 | 2 | 12 | 7 |
| G010.62 † | 0.53×0.41, -78 | 1.66e-04 | - | 46 | 5 | 32 | 9 |
| G012.80 | 1.48×1.48, 0 | 6.24e-04 | 3 | 37 | 3 | 25 | 9 |
| G351.77 † | 0.90×0.67, 87 | 4.39e-04 | - | 22 | 1 | 14 | 7 |
| G353.41 † | 0.95×0.67, 85 | 1.16e-03 | - | 20 | 2 | 11 | 7 |
| W43-MM1 | 0.56×0.56, 0 | 1.70e-04 | 3 | 56 | 3 | 43 | 9 |
| W43-MM2 | 0.54×0.54, 0 | 6.97e-05 | 3 | 29 | 3 | 16 | 10 |
| W43-MM3 | 0.52×0.52, 0 | 2.59e-05 | 3 | 34 | 1 | 25 | 8 |
| W51-E | 0.35×0.35, 0 | 3.14e-04 | 5 | 48 | 4 | 31 | 13 |
| W51-IRS2 | 0.51×0.51, 0 | 6.91e-05 | 7 | 81 | 6 | 69 | 6 |
| MGPS90 + ALMA-IMF-B3 | | | | | | | |
| G012.80 | 1.48×1.48, 0 | 1.45e-04 | 5 | 32 | 1 | 26 | 5 |
| W43-MM1 | 0.56×0.56, 0 | 2.50e-05 | 3 | 41 | 2 | 33 | 6 |
| W43-MM2 | 0.54×0.54, 0 | 2.13e-05 | 3 | 29 | 3 | 10 | 16 |
| W43-MM3 | 0.52×0.52, 0 | 3.00e-05 | 3 | 8 | 1 | 5 | 2 |
| W51-E | 0.35×0.35, 0 | 3.91e-05 | 3 | 81 | 3 | 66 | 12 |
| W51-IRS2 | 0.51×0.51, 0 | 9.70e-05 | 3 | 40 | 2 | 29 | 9 |

**Table 5.** Parameters and results for the dendrogram runs. Fields marked with † are available only in Band 6. *Beam* is the FWHM and PA of the beam of the map used for the dendrogram run, i.e., after convolution to a common beam when the spectral index map can be calculated. MAD$^{-\rho}$ is the noise estimate from the recursive MAD measured on the maps, $N_{\rm thr} \times$ MAD$^{-\rho}$ is the cut-off level used to calculate the spectral index. *Trunks* are the most extended structures, *leaves* are the most compact structures inside trunks. *Isolated* are structures without further hierarchical structure.

| Field | Leaves | Leaves$_{<0.1pc}$ | Leaves$_{\geq0.1pc}$ | Iso | Iso$_{<0.1pc}$ | Iso$_{\geq0.1pc}$ | Trunks | Trunks$_{<0.1pc}$ | Trunks$_{\geq0.1pc}$ |
|---|---|---|---|---|---|---|---|---|---|
| BGPS + ALMA-IMF-B6 | | | | | | | | | |
| G008.67-B6 † | 12 | 12 (100.0%) | 0 (0.0%) | 7 | 6 (85.7%) | 1 (14.3%) | 2 | 1 (50.0%) | 1 (50.0%) |
| G010.62-B6 † | 32 | 30 (93.7%) | 2 (6.3%) | 9 | 8 (88.9%) | 1 (11.1%) | 5 | 3 (60.0%) | 2 (40.0%) |
| G012.80-B6 | 24 | 23 (95.8%) | 1 (4.2%) | 9 | 9 (100.0%) | 0 (0.0%) | 3 | 0 (0.0%) | 3 (100.0%) |
| G351.77-B6 † | 14 | 14 (100.0%) | 0 (0.0%) | 7 | 7 (100.0%) | 0 (0.0%) | 1 | 0 (0.0%) | 1 (100.0%) |
| G353.41-B6 † | 11 | 11 (100.0%) | 0 (0.0%) | 7 | 7 (100.0%) | 0 (0.0%) | 2 | 0 (0.0%) | 2 (100.0%) |
| W43-MM1-B6 | 40 | 40 (100.0%) | 0 (0.0%) | 9 | 9 (100.0%) | 0 (0.0%) | 4 | 2 (50.0%) | 2 (50.0%) |
| W43-MM2-B6 | 12 | 12 (100.0%) | 0 (0.0%) | 10 | 9 (90.0%) | 1 (10.0%) | 3 | 1 (33.3%) | 2 (66.7%) |
| W43-MM3-B6 | 14 | 12 (85.7%) | 2 (14.3%) | 8 | 8 (100.0%) | 0 (0.0%) | 1 | 0 (0.0%) | 1 (100.0%) |
| W51-E-B6 | 31 | 30 (96.8%) | 1 (3.2%) | 13 | 13 (100.0%) | 0 (0.0%) | 4 | 1 (25.0%) | 3 (75.0%) |
| W51-IRS2-B6 | 59 | 54 (91.5%) | 5 (8.5%) | 7 | 7 (100.0%) | 0 (0.0%) | 5 | 0 (0.0%) | 5 (100.0%) |
| MGPS90 + ALMA-IMF-B3 | | | | | | | | | |
| G012.80-B3 | 26 | 23 (88.5%) | 3 (11.5%) | 3 | 3 (100.0%) | 0 (0.0%) | 1 | 0 (0.0%) | 1 (100.0%) |
| W43-MM1-B3 | 33 | 31 (94.0%) | 2 (6.0%) | 4 | 4 (100.0%) | 0 (0.0%) | 2 | 0 (0.0%) | 2 (100.0%) |
| W43-MM2-B3 | 7 | 7 (100.0%) | 0 (0.0%) | 8 | 8 (100.0%) | 0 (0.0%) | 2 | 1 (50.0%) | 1 (50.0%) |
| W43-MM3-B3 | 5 | 5 (100.0%) | 0 (0.0%) | 2 | 2 (100.0%) | 0 (0.0%) | 1 | 0 (0.0%) | 1 (100.0%) |
| W51-E-B3 | 45 | 45 (100.0%) | 0 (0.0%) | 5 | 5 (100.0%) | 0 (0.0%) | 2 | 0 (0.0%) | 2 (100.0%) |
| W51-IRS2-B3 | 22 | 20 (91.0%) | 2 (9.0%) | 7 | 6 (85.7%) | 1 (14.3%) | 2 | 0 (0.0%) | 2 (100.0%) |

**Table 6.** Number of leaves, trunks, and isolated structures in each field. The respective number and percentage of structures with sizes < 0.1 pc and ≥ 0.1 pc are given.

where $I$ is the intensity and $\nu$ is the frequency.

The radio and millimeter spectral index of thermal free-free emission varies from $\alpha_{\rm ff} = 2$ to $-0.1$ with increasing frequency, since its optical depth decreases with frequency. In contrast, for thermal dust emission, $\alpha_{\rm dust}$ varies from $2 + \beta$ to 2 as frequency increases, since the optical depth of this emission mechanism increases with frequency ($\beta \approx 1.5 - 2$ is the dust opacity index). At the frequencies of our continuum images free-free is expected to have small optical depths, whereas the optical depth of dust emission will range from negligible to moderate for the densest structures. Therefore, in our maps we define free-free emission to be dominant if $\alpha \leq 0.5$, and dust emission to dominate if $\alpha \geq 1.5$. We define the emission to be of mixed type if $0.5 < \alpha < 1.5$, i.e, with unknown relative contributions of free-free and dust.

For our calculations, we only consider pixels above a primary beam response $\geq 0.4$[8], and an intensity cut-off $\geq N_{\mathrm{thr}} \times \mathrm{MAD}^{-\rho}$ (see Table 5). The $N_{\mathrm{thr}}$ values were optimized from visual inspection in the range from 3 to 5 in order to be consistent with the structures identified by `astrodendro`. Only W51-IRS2 required a higher $N_{\mathrm{thr}}$ in Band 6 (see Table 5). Figure 7 presents the spectral index and emission error maps for the above mentioned fields. Figure 8 presents colored masks with the dominant emission type, as well as the spatial distribution of leaves, trunks, and isolated structures for W43-MM1, W43-MM2, W43-MM3, G012.80, W51-E, and W51-IRS2. We note that some areas in the spectral index maps are smaller than a resolution element because of the intensity cutoff requirement. FITS files with the spectral index maps and errors are distributed with this publication (see Appendix E). In this work we combine the information from the spectral index maps and dendrograms to infer the basic physical properties of the identified hierarchy of structures, taking into account their dominant emission process.

We compared the spectral index maps from our combination with those obtained with the purely interferometric data. As expected, they are consistent, but the spectral index is recovered over larger areas in the combined maps. Appendix B summarizes this comparison.

### 4.3. Dust contribution

For all fields the free-free emission contribution is significantly smaller at 1 mm than at 3 mm. Therefore, we use the Band 6 dendrograms in the combined images as the basis for calculating the thermal dust emission and for estimating both the dust mass and column density for each field and structure. For each field, we extract the dendrogram hierarchy and then we consider the emission-type information on a pixel-by-pixel basis using the spectral index masks. Finally, we get the fluxes integrated over the respective structure using only those pixels labeled as dust.

---

[8] The FOV of the Band 3 mosaics is slightly larger than for Band 6 due to the larger single-pointing primary beam at longer wavelengths (see Ginsburg et al. 2022), which means that the spectral index maps are limited by the Band 6 FOV.

To estimate the dust mass, we consider the radiative transfer equation neglecting background and with a constant source function:

$$I_\nu = B_\nu [1 - \mathrm{e}^{-\tau_\nu}], \qquad (3)$$

where $I_\nu$ is the pixel intensity or the average intensity of the dust emission in the dendrogram structure, depending on whether we are calculating a column density map or the average properties of a structure. $B_\nu$ is the Planck function at the dust temperature $T_{\mathrm{dust}}$. We then use:

$$\Sigma_{\mathrm{dust}} = \frac{\tau_\nu}{\kappa_\nu}, \qquad (4)$$

where $\Sigma_{\mathrm{dust}}$ is the dust surface density in g cm$^{-2}$ and $\kappa_\nu$ is the dust opacity. We set $\kappa_{1.3\mathrm{mm}} = 1$ cm$^2$ g$^{-1}$ (Ossenkopf & Henning 1994), as used in previous ALMA-IMF studies (e.g., Pouteau et al. 2023; Nony et al. 2023). We use $T_{\mathrm{dust}}$ maps at $10''$ resolution determined applying the PPMAP technique to the far-IR and millimeter photometry (Marsh et al. 2017). The PPMAP results for the ALMA-IMF sample will be presented in Dell'Ova et al. (in prep.). We assume a 10% error in the $T_{\mathrm{dust}}$ determination.

Finally, the optical depth $\tau_\nu$ and dust mass $M_{\mathrm{dust}}$ are calculated using:

$$\tau_\nu = -\ln\left(1 - \frac{I_\nu}{B_\nu}\right), \qquad (5)$$

$$M_{\mathrm{dust}} = \frac{\Omega d^2}{\kappa_\nu} \tau_\nu, \qquad (6)$$

where $\Omega$ is the angular size of the emission (pixel or dendrogram structure) and $d$ is the distance to the source. The previous equations intentionally avoid making the two common assumptions of: $i$) being in the Rayleigh-Jeans regime (R-J, $h\nu \ll k_{\mathrm{B}}T$), and $ii$) having optically-thin emission ($\tau_\nu \ll 1$). We have verified that small optical depths are the case for most pixels, but we note that assuming $i$) could lead to overestimate masses from a few to more than a hundred percent. The reason for this is that, at a given frequency, the bias introduced by the R-J assumption increases at small optical depths. This effect is more relevant at the relatively high frequencies of ALMA observations.

The derived quantities of the dust-dominated dendrogram structures are listed in Table 9 in Appendix D. Some structures are present in both the W51-IRS2 and W51-E mosaics. We avoid this overlap by excluding them from the list of W51-IRS2 structures. This affects five structures in total. To derive the molecular gas column density maps for the pixels dominated by dust emission, we assume a gas-to-dust mass ratio $M_{\mathrm{gas}}/M_{\mathrm{dust}} = 100$, and use the equation:

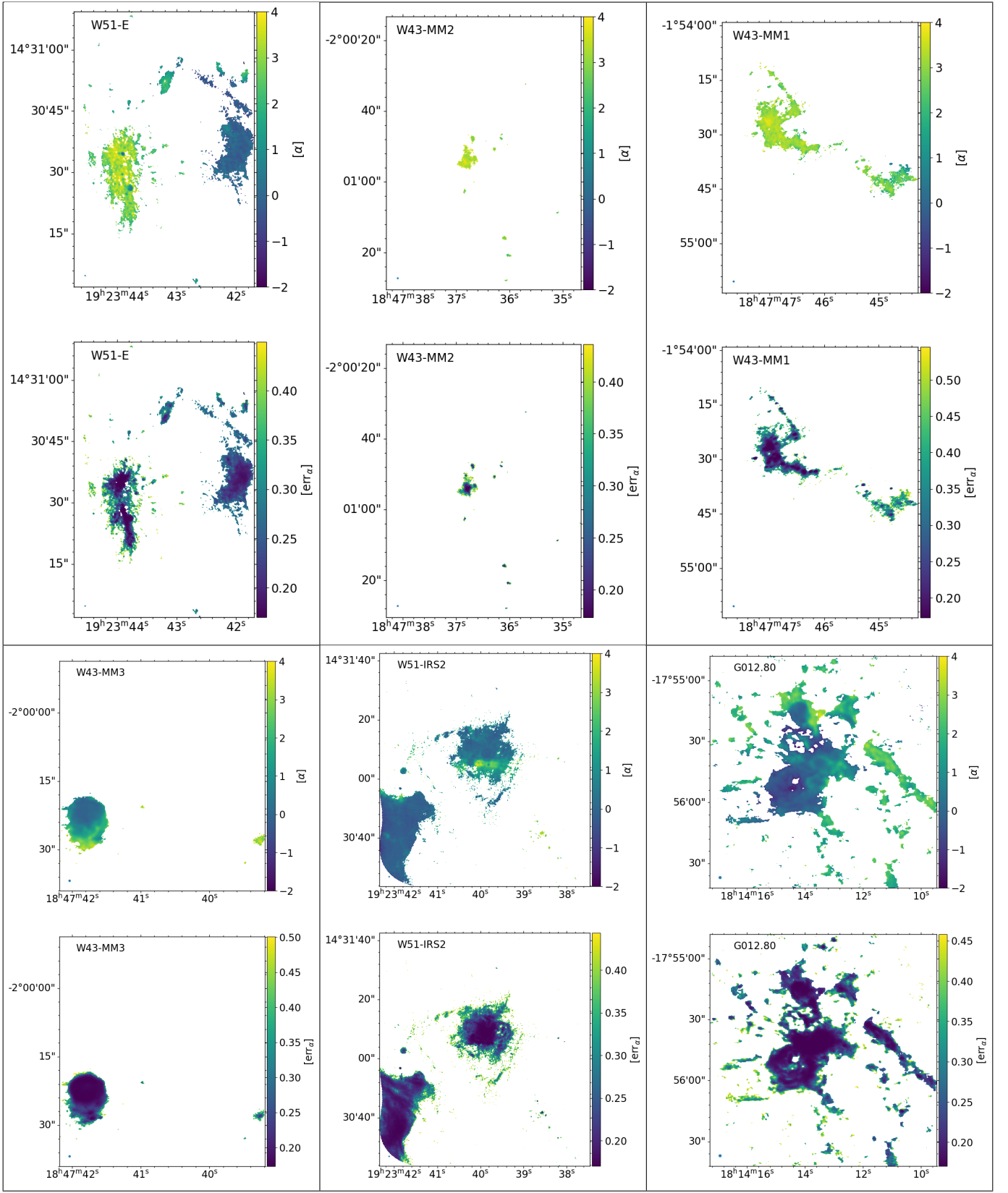

**Figure 7.** For each field the top sub-panel presents the Band 3 to Band 6 spectral index map, and the bottom sub-panel presents the corresponding error map.

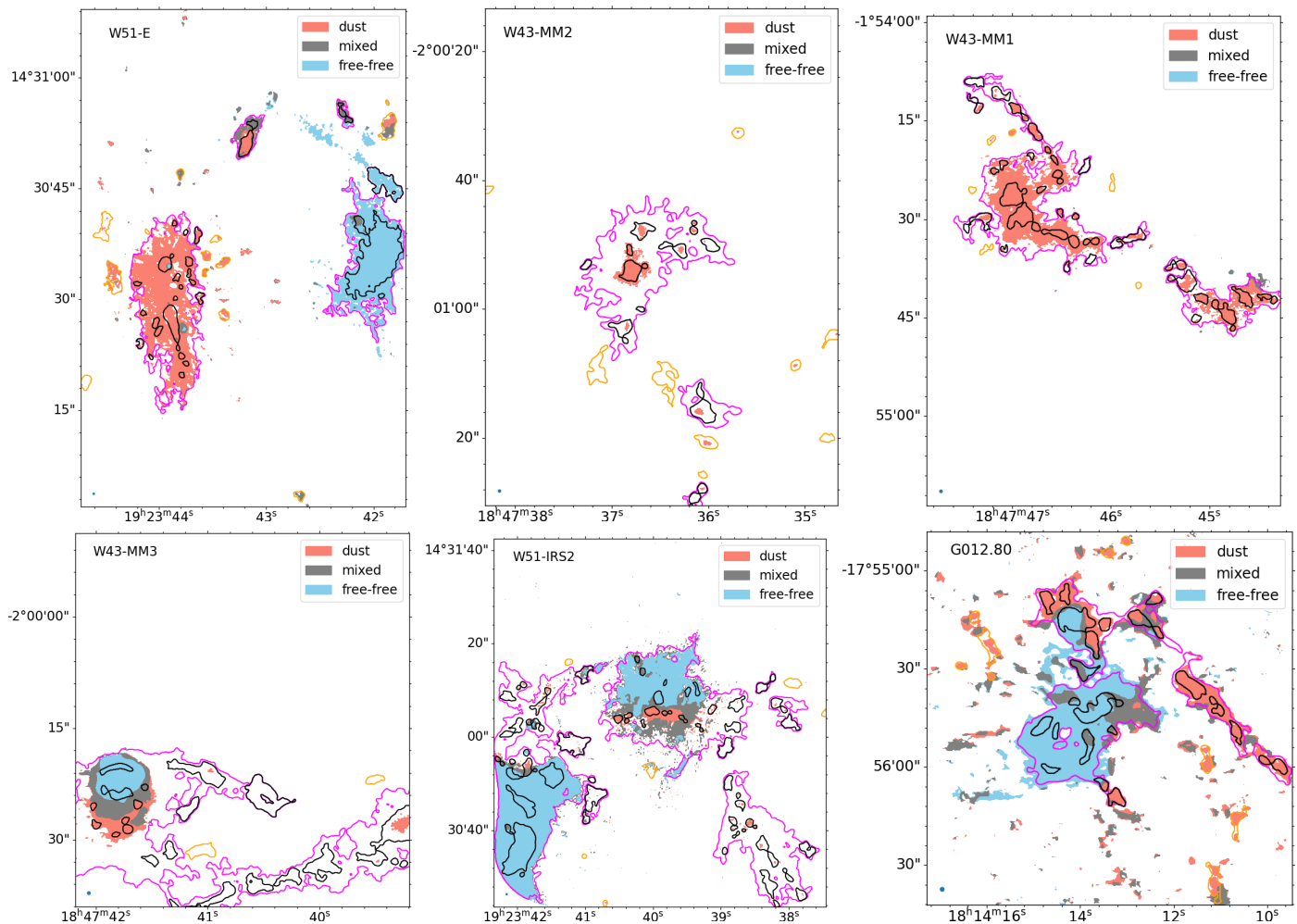

**Figure 8.** Colored pixels show the dominant emission type based on the Band 3 to Band 6 spectral index. $\alpha \geq 1.5$ is labeled as dust emission (salmon), and $\alpha \leq 0.5$ corresponds to free-free emission (light blue). Intermediate spectral indices are labeled as mixed (gray). Overlaid contours show the spatial distribution of leaves (black), trunks (magenta), and isolated structures (orange), measured in the Band 6 images.

$$N_{\text{gas}} = \frac{M_{\text{gas}}}{M_{\text{dust}}} \times \frac{M_{\text{dust}}}{\Omega m_{\text{H}} \mu},\qquad(7)$$

where $\Omega$ is the solid angle, $\mu = 2.8$ is the mean molecular weight (Kauffmann et al. 2008), and $m_{\text{H}}$ is the mass of a hydrogen atom. Figure 14 in Appendix C presents the column density maps for each of the six regions with both Band 3 and Band 6 data combination.

### 4.4. Free-free contribution

Estimations of the free-free emission are useful not only to remove contamination from the dust emission, but also to characterize the HII emission in the ALMA-IMF fields. Since the free-free emission may be relatively strong compared to dust at 3 mm, we use the Band 3 dendrograms together with the spectral index maps to locate it and to estimate the emission measure (EM) and electron densities ($n_e$) for each dendrogram structure with pixels labeled as free-free. For each field we extract the dendrogram hierarchy and then use the emission type information from the spectral index masks as to obtain the integrated fluxes from the pixels labeled as free-free emission.

We use the measured band 3 fluxes and sizes to derive the emission measure EM in pc cm$^{-6}$ and $n_e$ in cm$^{-3}$ as in Rivera-Soto et al. (2020)[9]:

$$\text{EM} = 12.143\tau_\nu \left(\frac{\nu}{\text{[GHz]}}\right)^{2.1} \left(\frac{T_e}{\text{[K]}}\right)^{1.35},\qquad(8)$$

$$n_e = \sqrt{\frac{\text{EM}}{D}}.\qquad(9)$$

We note that Equation 8 assumes optically-thin free-free emission. This is always valid at millimeter wavelengths, as we have verified from the small measured brightness temperatures. We also use Equation 5 in its full Planck form at an electron temperature $T_e = 8000$ K. We define the effective diameter $D$ in pc as $D = 2\sqrt{A/\pi}$, where $A$ is the area of the pixels labeled as free-free in the dendrogram structure. Equation 9 for $n_e$ assumes circular geometry in the plane of the sky. We have measured the ellipticity[10] of the detected structures. In the Band 3 dendrograms, all fields have mean $e$ values ranging from 0.20 to 0.51. In the Band 6 dendrograms, the mean $e$ ranges from 0.30 to 0.68. Therefore, the assumption of spherical geometry is reasonable, although not ideal.

Finally, we determine the number of Lyman continuum photons per second $Q_0$ required to produce the observed free-free emission, assuming ionization-recombination balance (Osterbrock & Ferland 2006):

$$Q_0 = \frac{\pi}{6}\alpha_{\text{B}} n_e^2 D^3,\qquad(10)$$

where $\alpha_{\text{B}} = 2 \times 10^{-13}$ cm$^3$ s$^{-1}$ is the case-B hydrogen recombination coefficient. The derived quantities of the HII structures are listed in Table 10 in Appendix D. As mentioned previously, to estimate the free-free contribution in W51-IRS2, we exclude four dendrogram structures that overlap with the W51-E mosaic.

The derived HII region electron densities are around $n_e \sim 10^4$ cm$^{-3}$, as expected for ultracompact (UC) HII regions (Hoare et al. 2007). Sizes range from $\sim 10^{-1}$ to $10^{-2}$ pc, as expected for UC and hypercompact (HC) HII regions. We note that several of these millimeter-detected objects have hypercompact sizes (0.01 pc) but relatively low densities ($10^4$ cm$^{-3}$). A similar finding was reported by Rivera-Soto et al. (2020) using centimeter data, and was interpreted as the detection of a more 'normal' population of very small HII regions. Rivera-Soto et al. (in prep.) will present a VLA catalog of UC and HC HII regions in the ALMA-IMF fields observable from the northern hemisphere.

## 5. DISCUSSION

### 5.1. Large to Small Scale Efficiencies

We investigated the efficiency of gas fragmentation between scales in the dendrogram hierarchy of dust-dominated structures, in order to find hierarchical relations and to evaluate their possible link to the evolution of the amount of ionization in the ALMA-IMF proto-clusters. Louvet et al. (2014) found an almost linear correlation between the core formation efficiency (CFE) and volumetric density in concentric areas of the W43-MM1 clump. Similarly, Csengeri et al. (2017) found a correlation between the CFE and the average density of a sample of massive star forming clumps.

We looked for analog relations in the dust emission using as proxy the fraction of molecular gas mass in leaves ($M_{\text{g,lv}}/M_{\text{g,tot}}$) within a threshold gas column density $N_{\text{gas}}$. The threshold levels are defined by each of the identified dendrogram branches and trunks. In this definition, $M_{\text{g,lv}}$ corresponds to the total gas mass in leaves in pixels labeled as dust-dominated within a contour defined by $N_{\text{gas}}$. Similarly, $M_{\text{g,tot}}$ corresponds to the total gas mass in dust-dominated pixels within the same contour, regardless of which level the dendrogram hierarchy they belong to. We note that with our approach we do not attempt to derive volume densities or star formation efficiencies (SFEs), yet it can offer a comparative view of several clouds using a homogeneous method. Converting the data in Table 3 of Louvet et al. (2014) to average column densities assuming circular geometry, their measurements of the CFE for W43-MM1 are in the range $N_{gas} \sim 5 \times 10^{22}$ to $9 \times 10^{23}$ cm$^{-2}$, consistent with our range of column densities.

---

[9] More details on the physical assumptions can be found in chapter 10 of Wilson et al. (2009).

[10] Defined as $e = 1 - b/a$, where $a$ and $b$ are the major and minor semiaxes, i.e., the ellipticity of a circle is zero.

As expected from its definition using the dendrogram hierarchy, the LME is a monotonically increasing function of $N_{gas}$ (see Figure 9). However, there are a few differences in the shape of the LME curves. All fields except G012.80 show an ever increasing or exponential-like growth in the logarithmic plots. G012.80 is also the only field that reaches a 100% LME at its highest $N_{gas}$, with signs of saturation before reaching its maximum value. In contrast, W51-E has the lowest peak LME of the sample at $\lesssim 50\%$, with no signs of saturation. Motte et al. (2022) give estimations of the percentage of mass in star-forming cores as compared to the total mass recovered in the interferometric ALMA-IMF images. From the six fields that we analyze, W51-E and G012.80 have the smallest (5%) and largest (32%) percentages of mass in cores, consistent with our measurements.

A preliminary interpretation of Figure 9 is that clumps with an exponentially increasing LME, i.e., those that form increasingly more substructures with increasing column densities, are currently in a very active period of star formation. From the six analyzed fields, only possibly G012.80 does not satisfy this criterion. The LME vs $N_{gas}$ curves in Figure 9 are color-coded with an estimate of the protocluster star formation rate (SFR$_{IR}$, see also Table 7 ) using the near-IR to millimeter luminosities from Dell'Ova (in prep.), and applying the extragalactic calibration of Murphy et al. (2011)[11]. W51-E and W51-IRS2 stand out as the most star-forming, and the only ones with peak $N_{gas}$ thresholds approaching $10^{25}$ cm$^{-2}$. However, extragalactic calibrations of the star formation rate are sensitive to stellar populations averaged over $\sim$ 10 to 100 Myr, and are inferred from large volumes where the stellar IMF can be considered as fully sampled (for a review, see Kennicutt & Evans 2012). None of these assumptions fully applies to individual molecular clouds (e.g., Jáquez-Domínguez et al. 2023), although it is possible that in the more evolved W51 regions the stellar IMF is already well populated (see below).

There is evidence that the W51 protoclusters are currently converting a significant fraction of their gas into stars (Kumar et al. 2004; Saral et al. 2017), despite the extensive stellar feedback that is already present (Ginsburg et al. 2016). As mentioned before, W51-IRS2 and W51-E are the two fields with the highest peak $N_{gas}$ in Figure 9. However, the maximum LME of W51-IRS2 surpasses 90%, whereas for W51-E the peak LME is $\lesssim 50\%$. This indicates that even at its highest column densities, a significant fraction of the molecular mass in W51-E is not in leaves, but in lower-level structures within its own dendrogram hierarchy. This can also be seen in Figures 8 and 14: in W51-E, most of the emission labeled as dust-dominated is above $N_{gas} > 10^{24}$ cm$^{-2}$, but the dendrogram leaves only cover a small fraction of this area. The majority of this high column-density dust emission is not close to the H II arc located about 0.8 pc to the west (see Fig. 8), thus it is unlikely to be the the product of triggering from the expansion of the H II region.

Regarding the W43 protoclusters (MM1, MM2, and MM3), it has been proposed that they are in a "mini-starburst" stage. This proposal is based on their copious formation of cores from low to high masses, as seen in a variety of metrics such as their CFE, estimates of their SFE, and measurements of the slope of their CMF (Louvet et al. 2014; Motte et al. 2018b; Pouteau et al. 2023). However, these protoclusters are significantly younger than W51-IRS2, W51-E, or G012.80. Therefore, the inferred SFR$_{IR}$ is probably an underestimation of the actual current star formation rate, mostly dominated by a large popoulation of pre- and protostellar cores (Nony et al. 2023).

Finally, G012.80 (also known as W33 Main) stands out because it has the highest peak LME (100%) of all fields, yet the final increments of LME vs $N_{gas}$ show signs of slowing down. G012.80 has a well known H II region and is the more evolved of the star forming clumps in the W33 complex (Immer et al. 2014; Khan et al. 2022). In contrast to W51-E, in Fig. 8 it is seen that the dendrogram leaves in G012.80 cover the majority of the dust-dominated area, and that they closely surround the prominent H II region from the north and west sides. It is possible that H II region feedback has somewhat reduced the total potential of G012.80 for subsequent star formation, while at the same time inducing significant fragmentation in the remaining material around the H II region (e.g., Dale et al. 2014; Jáquez-Domínguez et al. 2023). The current star formation activity of G012.80 is being studied in detail by Armante et al. (submitted). Given its evolved stage, our estimatate of SFR$_{IR}$ might be as good for G012.80 as it is for the W51 protoclusters. Therefore, it appears that the star formation activity in G012.80 might indeed be past its peak.

[11] $\left(\frac{\text{SFR}_{IR}}{M_\odot \text{ yr}^{-1}}\right) = 3.88 \times 10^{-44} \left(\frac{L_{IR}}{\text{erg s}^{-1}}\right)$.

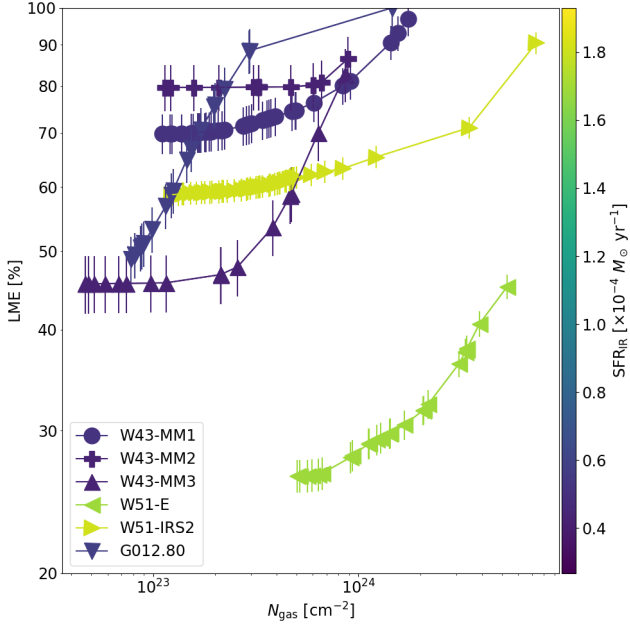

**Figure 9.** Leaf Mass Efficiency (LME) as a function of $N_{\rm gas}$, the threshold molecular gas column density of the dendrogram level (branch or trunk) at which the LME is measured. Errors in the LME are propagated analytically from the gas mass estimations. The color scheme refers to estimates of the protocluster star formation rate ($\rm SFR_{IR}$) based on their infrared luminosity (see text).

Interestingly, within the ALMA-IMF sample, the power-law index of the CMF of evolved regions such as G012.80 appears to be closer to the Salpeter slope compared to younger regions (Louvet et al. submitted). In other words, the most evolved regions do not present the relative excess of massive cores that younger regions such as the W43 fields have (Motte et al. 2018b; Pouteau et al. 2022; Nony et al. 2023). Our tentative interpretation is that the behaviour of the LME vs $N_{\rm gas}$ curve could be physically related to the CFE slope, and that both of them could be ultimately linked to the effects of feedback during cloud evolution. Further measurements and interpretative work are warranted.

### 5.2. *Evolutionary stages*

The amount of ionizing feedback is a potential way of tracing the evolution of massive star formation regions. This is usually done using as tracers the free-free continuum and hydrogen recombination line emission associated to compact and ultracompact (UC) H II regions (e.g., Kurtz et al. 1994; Kalcheva et al. 2018; Rivera-Soto et al. 2020). For the ALMA-IMF sample, Motte et al. (2022) proposed an evolutionary sequence based on the far-IR emission and the surface density of H41$\alpha$ recombination line emission. In this paper, us-ing the continuum data combination, we have looked for regions dominated by free-free in the 3 mm maps of the six fields with spectral index information (see Section 4.4). Four fields (G012.80, W43-MM3, W51-E, and W51-IRS2) have detectable free-free structures in the dendrogram identification. We then estimated the basic physical properties of their H II regions, including the ionizing photon rate $Q_0$ (see Table 10).

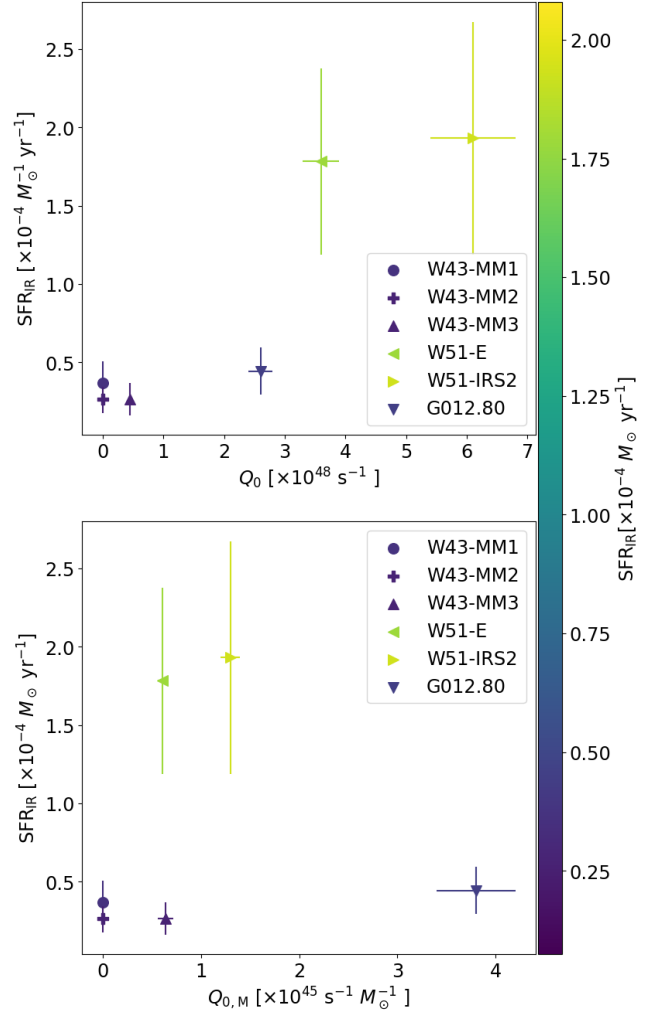

**Figure 10.** Estimates of $\rm SFR_{IR}$ as a function of $Q_0$ (top panel) and $Q_{0,\rm M}$ (bottom panel).

| Field | $\rm SFR_{IR}$ [$\times 10^{-5}\ M_\odot\ \rm yr^{-1}$] | $Q_0$ [$\times 10^{48}\ \rm s^{-1}$] | $Q_{0,\rm M}$ [$\times 10^{45}\ \rm s^{-1}\ M_\odot^{-1}$] |
|---|---|---|---|
| W43-MM1 | $3.7 \pm 1.3$ | – | – |
| W43-MM2 | $2.7 \pm 0.9$ | – | – |
| W43-MM3 | $2.7 \pm 1.0$ | $0.45 \pm 0.04$ | $0.64 \pm 0.08$ |
| W51-E | $17.8 \pm 5.9$ | $3.6 \pm 0.3$ | $0.61 \pm 0.05$ |
| W51-IRS2 | $19.3 \pm 7.4$ | $6.1 \pm 0.7$ | $1.3 \pm 0.1$ |
| G012.80 | $4.5 \pm 1.5$ | $2.6 \pm 0.2$ | $3.8 \pm 0.4$ |

**Table 7.** Evolutionary proxies. Only four fields have measurements of photoionization rates using free-free emission. $\rm SFR_{IR}$ is likely a large underestimation of the actual star formation rate (see text). Measurements use the infrared to millimeter luminosities from Dell'Ova (in prep.), applying the calibration of Murphy et al. (2011).

We explore a possible evolutionary trend among the four fields with measured free-free emission. We use two evolution proxies: the total rate of ionizing photons $Q_0$ (s$^{-1}$) measured in the maps; and the ratio of $Q_0$ with the total mass of molecular gas, which we label as $Q_{0,\mathrm{M}} = Q_0/M_{\mathrm{g,tot}}$ (s$^{-1}$ $M_\odot^{-1}$). A quantity such as $Q_{0,\mathrm{M}}$ is analogous to the luminosity to mass ratio $L/M$, which is a widely used proxy for the evolution of star formation from protostellar to cloud scales (e.g., Molinari et al. 2008; Giannini et al. 2012; Giannetti et al. 2013; Elia et al. 2017), with the important difference that $Q_{0,\mathrm{M}}$ is only sensitive to the massive stellar population capable of producing hydrogen ionizing photons. Table 7 lists the evolutionary proxies for the four regions with measurements. These are consistent with the evolutionary scheme outlined in Motte et al. (2022) (see their Table 4), in which W43-MM3 is significantly younger than the other three fields. W51-IRS has the largest $Q_0$, but lies behind G012.80 in terms of $Q_{0,\mathrm{M}}$, which would be the most evolved field in terms of the $Q_{0,\mathrm{M}}$ ratio.

We look for relations between the estimates for the star formation rate of the analyzed regions and the proposed evolutionary proxies $Q_0$ and $Q_{0,\mathrm{M}}$ (see Figure 10). G012.80 clearly stands out as evolved, yet with a relatively low SFR$_{\mathrm{IR}}$. In contrast, the also relatively evolved W51-E and W51-IRS2 protoclusters have an order of magnitude larger SFR$_{\mathrm{IR}}$. At first glance, this result is consistent with our hypothesis that feedback from H II regions has decelerated the star formation process in G012.80, but has not had a disruptive effect in W51 (see also Ginsburg et al. 2016). However, it is also possible that the W51 protoclusters are more massive and luminous in absolute terms, i.e., that other protoclusters within our sample will not reach, or never reached, the sheers amounts of star formation in W51.

## 6. SUMMARY AND CONCLUSIONS

With this paper we provide a release of a combination of the ALMA-IMF continuum maps with single-dish bolometer surveys. The combination was performed using the feathering technique, which offers the compromise of being in Fourier-space, yet computationally inexpensive compared to joint deconvolution. The input for combination are the data products labeled as *bsensnobright* in the ALMA-IMF continuum image release (Ginsburg et al. 2022), as well as images from the Mustang-2 Galactic Plane Survey pilot at 3 millimeters (MGPS90, Ginsburg et al. 2020) and the Bolocam Galactic Plane Survey (BGPS, Aguirre et al. 2011; Ginsburg et al. 2013). Six and ten out of the fifteen ALMA-IMF fields are available for combination in Band 3 and Band 6, respectively.

We ran `astrodendro` (Robitaille et al. 2019) on the combined images and created spectral-index maps in homogenized versions of the six fields with combination in both bands (G012.80, W43-MM1, W43-MM2, W43-MM3, W51-IRS2, and W51-E). We then used the spectral index information to separate areas in the maps dominated by dust and free-free emission. Using the dendrogram structure identification, we estimated the basic physical properties of the dust-dominated (column densities and masses) and free-free dominated (emission measures, ionizing-photon rates) structures. We also calculated pixelized column-density maps of the dust-dominated areas.

With the previous measurements, we looked for multi-scale relations between the leaves (with sizes < 0.1 pc) and their corresponding branches and trunks (with sizes typically between 0.1 to 1 pc) in the dendrogram hierarchy. We defined a Leaf Mass Efficiency (LME) as the fraction of mass in dendrogram leaves, and explored its dependence with the corresponding threshold molecular gas column density $N_{\mathrm{gas}}$ in their parent structures. A variety of maximum LME values and behaviors of the LME vs $N_{\mathrm{gas}}$ curves are observed. We hypothesized that a rapidly increasing LME with $N_{\mathrm{gas}}$ attests to the current high star formation activity of the analyzed protoclusters. In the stand out cases of G012.80 and W51-E, the respectively high (100%) and low (≲ 50%) peak LMEs, as well as a slowing down of the LME curve in G012.80, are interpreted as signaling that star formation is just past its peak in the former, but still building up in the latter despite the presence of significant ionizing feedback. This interpretation is consistent with estimates of their current star formation rates based on infrared to millimeter photometry.

We also looked for evolutionary trends in the ionizing photon rate $Q_0$ and its normalization to the current amount of molecular gas $Q_0/M_{\mathrm{g}}$. These quantities could only be measured in the four fields W43-MM3, W51-IRS2, W51-E, G012.80 (in the two younger W43 fields they are consistent with zero). The results are in accordance with the evolutionary classification proposed by Motte et al. (2022) for the entire ALMA-IMF sample.

The authors acknowledge the anonymous referee for an encouraging and useful report that helped us to improve this paper. This paper makes use of the following ALMA data: ADS/JAO.ALMA#2017.1.01355.L, #2013.1.01365.S, and #2015.1.01273.S. ALMA is a partnership of ESO (representing its member states), NSF (USA) and NINS (Japan), together with NRC (Canada), MOST and ASIAA (Taiwan), and KASI (Republic of Korea), in cooperation with the Republic of Chile. The Joint ALMA Observatory is operated by ESO, AUI/NRAO and NAOJ. The project leading to this publication has received support from ORP, that is funded by the European Union's Horizon 2020 research and innovation programme under grant agreement No 101004719 [ORP]. DDG, RGM, and RRS acknowledge support from UNAM-PAPIIT project IN108822, and from CONACyT Ciencia de Frontera project ID: 86372. RGM also acknowledges support from the AAS Chrétien International Research Grant. AS gratefully acknowledges support by the Fondecyt Regular (project code 1220610), and ANID BASAL projects ACE210002 and FB210003. F.M., P.DO., N.C., F.L, M.A., and A.Gu. acknowledge support from the French Agence Nationale de la Recherche (ANR) through the project COSMHIC (ANR-20-CE31-0009). F.M., N.C., and F.L. also acknowledge support from the European Research Council (ERC) via the ERC Synergy Grant ECO-GAL (grant 855130). PS was partially supported by a Grant-in-Aid for Scientific Research (KAKENHI Number JP22H01271 and JP23H01221) of JSPS. R. A. gratefully acknowledges support from ANID Beca Doctorado Nacional 21200897. This research made use of astrodendro, a Python package to compute dendrograms of Astronomical data (http://www.dendrograms.org).

*Software:* astropy (Astropy Collaboration et al. 2013, 2018, 2022), uvcombine (Koch & Ginsburg 2022), astrodendro (Robitaille et al. 2019)

## APPENDIX

### A. COMBINED IMAGES

Figures 11 and 12 show the interferometric, single-dish, and combined images at 3 mm (6 fields) and 1 mm (10 fields), respectively.

### B. COMPARISON OF SPECTRAL INDICES

Table 8 shows the areas with a valid spectral index measurement ($5\sigma$ noise threshold) before and after combination. Figure 13 shows the difference between the spectral Band 3 to Band 6 spectral index $\alpha$ using the combined and interferometric data sets. As expected, the values are consistent, with differences increasing toward the edges where extended emission is more dominant.

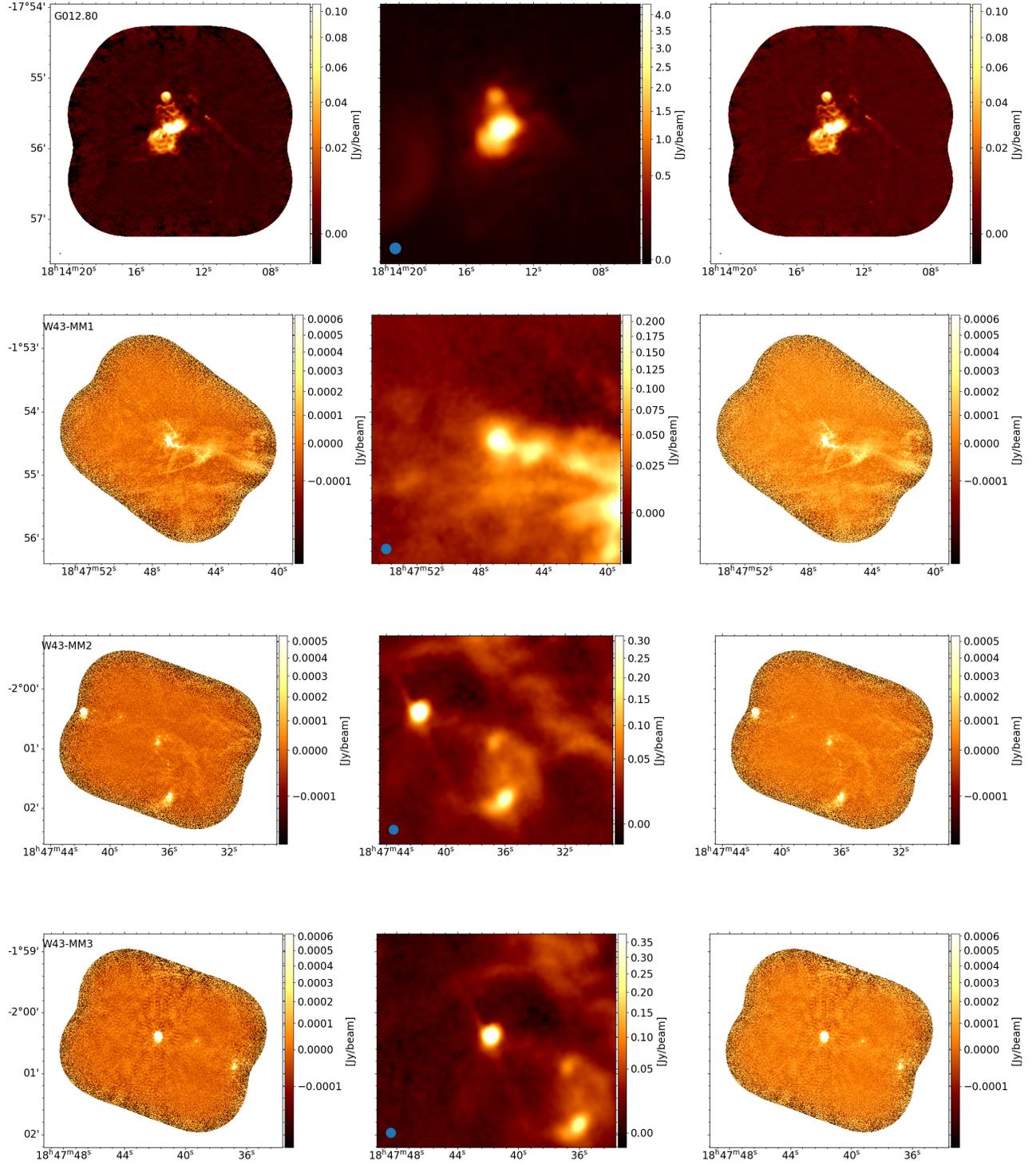

**Figure 11.** For each field, the left sub-panel presents the ALMA-IMF Band 3 data, the middle sub-panel shows the MGPS90 data, and the right sub-panel shows the combined data. The color scale in the middle column goes from $-3 \times \mathrm{MAD}^{-\rho}$ to the maximum value in the map, using a square-root color mapping. For the interferometric and combined images the color mapping follows the same rule, but the minimum and maximum between the two images is used for both plots.

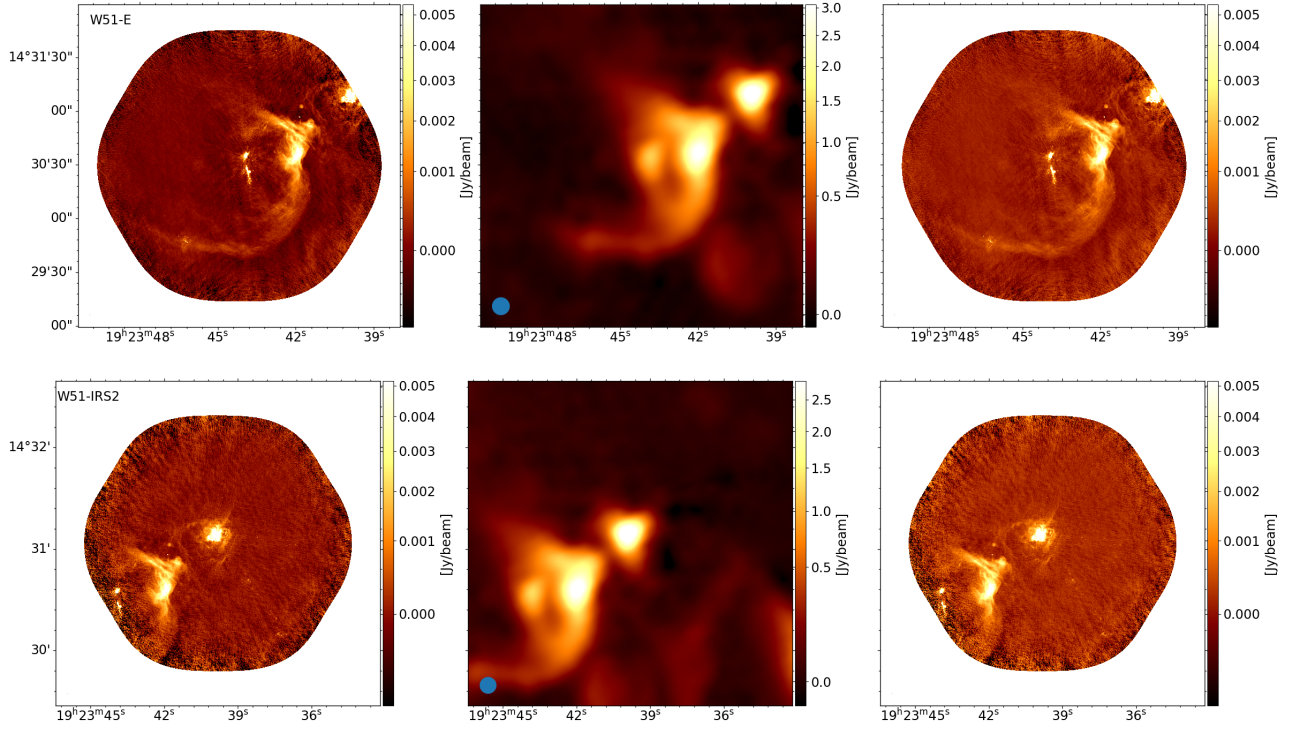

**Figure 11.** Continued. From top to bottom : W51-E, W51-IRS2.

| Field | $A_{\mathrm{int}}$ [$''2$] | $A_{\mathrm{comb}}$ [$''2$] | $A_{\mathrm{comb}}/A_{\mathrm{int}}$ |
|---|---|---|---|
| G012.80 | 1033.23 | 2468.46 | 2.39 |
| W43-MM1 | 129.23 | 184.32 | 1.43 |
| W43-MM2 | 18.54 | 25.75 | 1.39 |
| W43-MM3 | 80.31 | 88.46 | 1.10 |
| W51-E | 143.88 | 287.88 | 2.00 |
| W51-IRS2 | 459.23 | 709.97 | 1.55 |

**Table 8.** $A_{\mathrm{int}}$ and $A_{\mathrm{comb}}$ are the areas where the spectral index maps are measured in ALMA-only and combined images, using the same procedure.

## C. COLUMN DENSITY MAPS

Figure 14 shows the column density maps for each region, computed using only pixels identified as dominated by dust in our spectral index maps. As a consequence, the resulting maps may contain a few patches smaller than the resolution element.

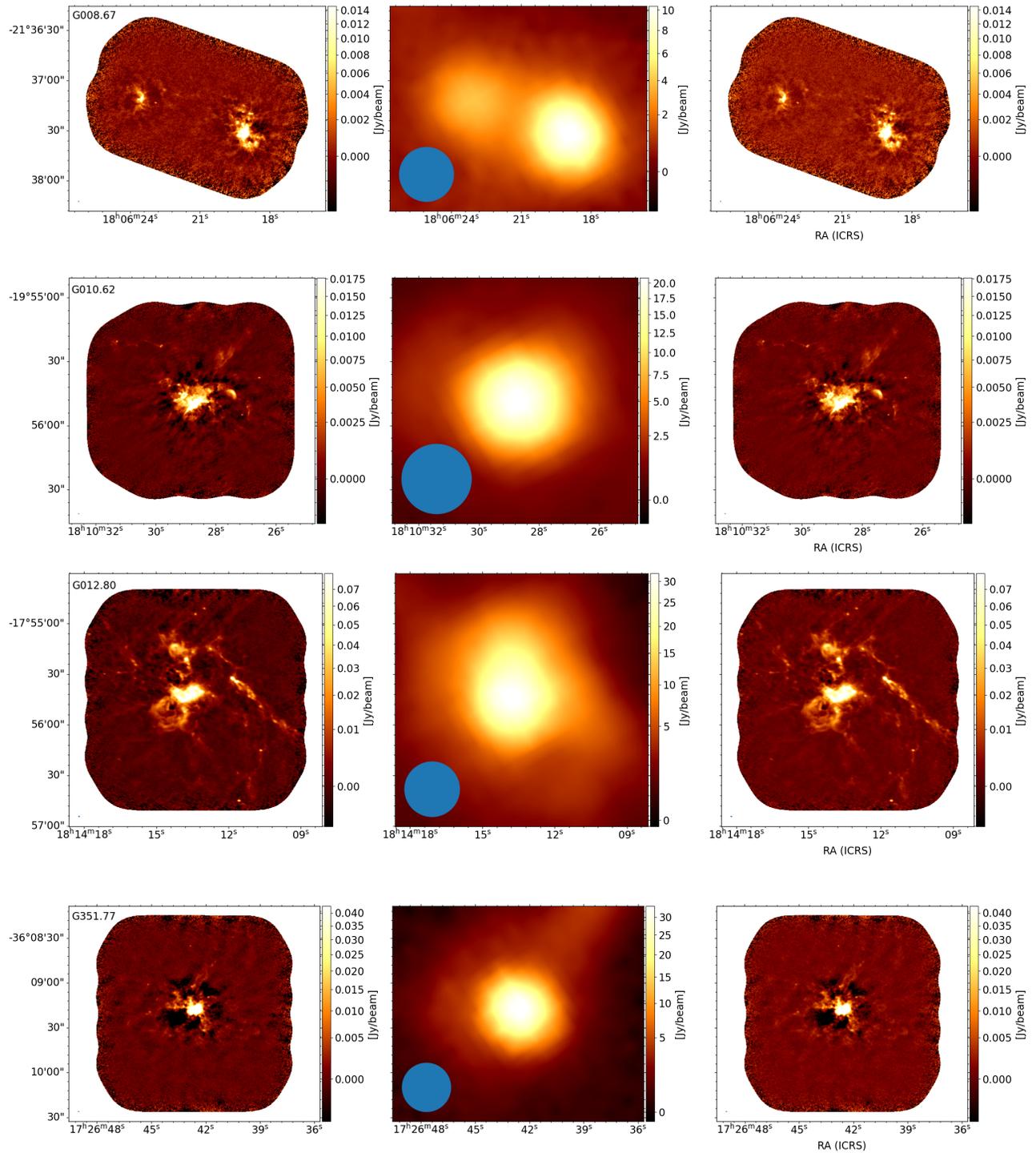

**Figure 12.** For each field the left sub-panel presents the ALMA-IMF Band 6 data, the middle sub-panel shows the BGPS data, and the right sub-panel shows the combined data. The color scale in the middle column goes from $-3 \times \mathrm{MAD}^{+\rho}$ to the maximum value in the map, using a square-root color mapping. For the interferometric and combined images the color mapping follows the same rule, but the minimum and maximum between the two images is used for both plots.

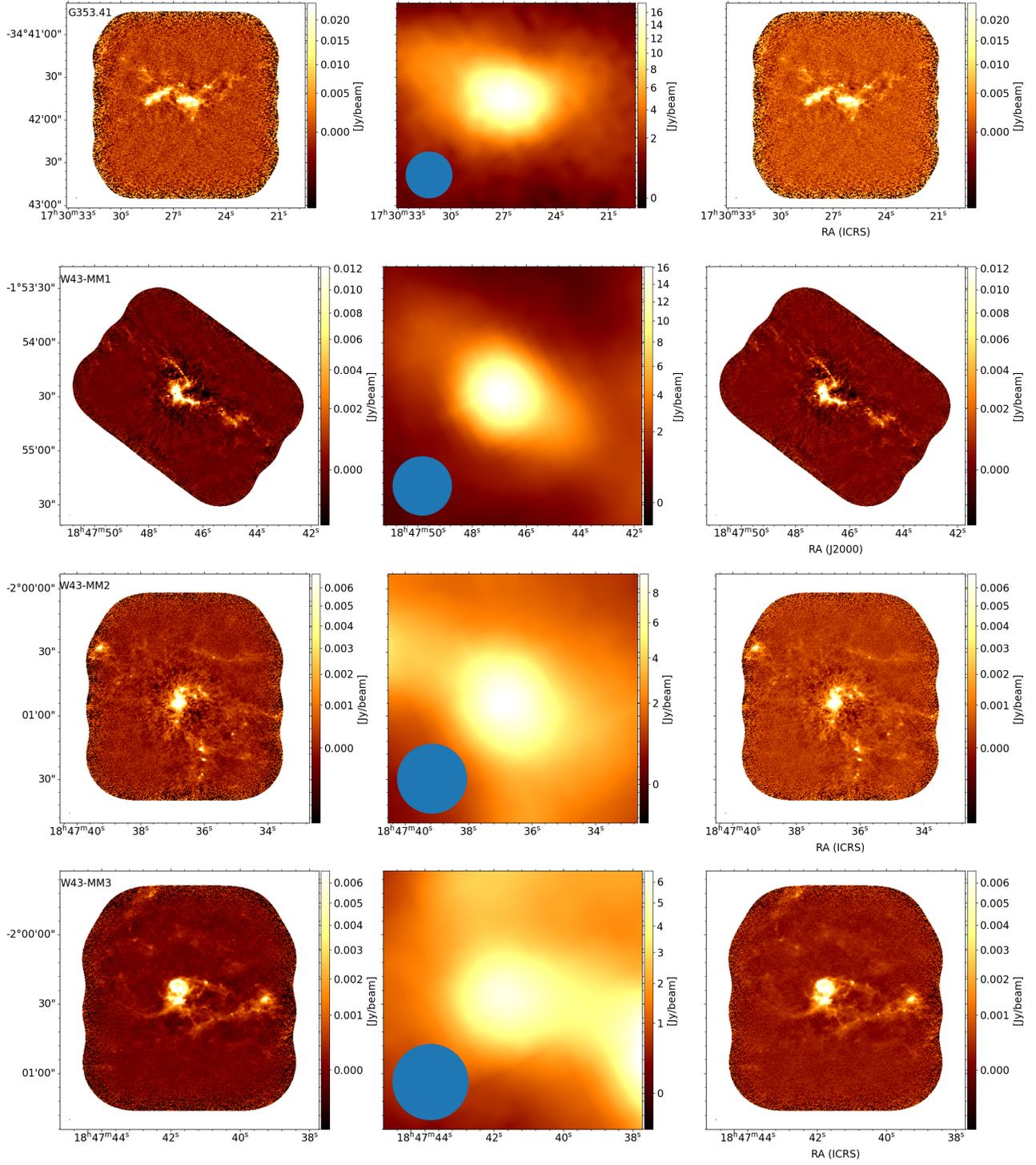

**Figure 12.** Continued.

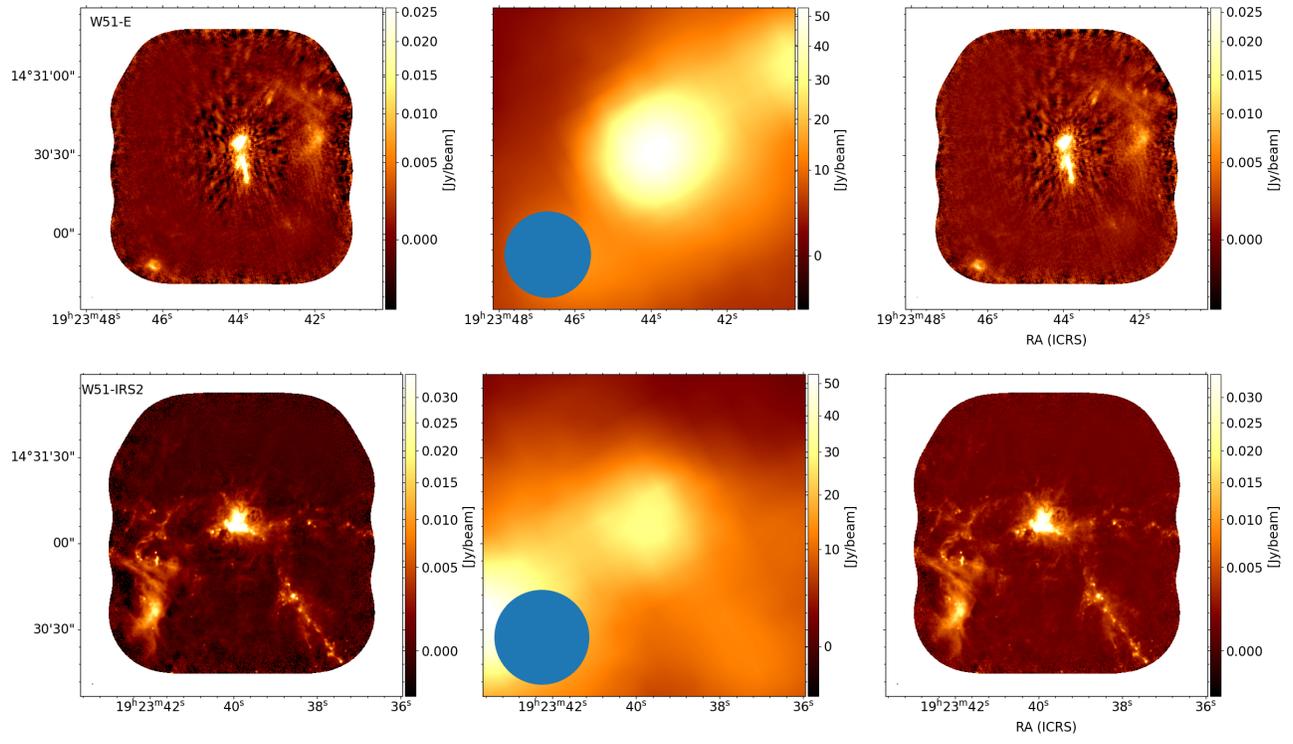

**Figure 12.** Continued.

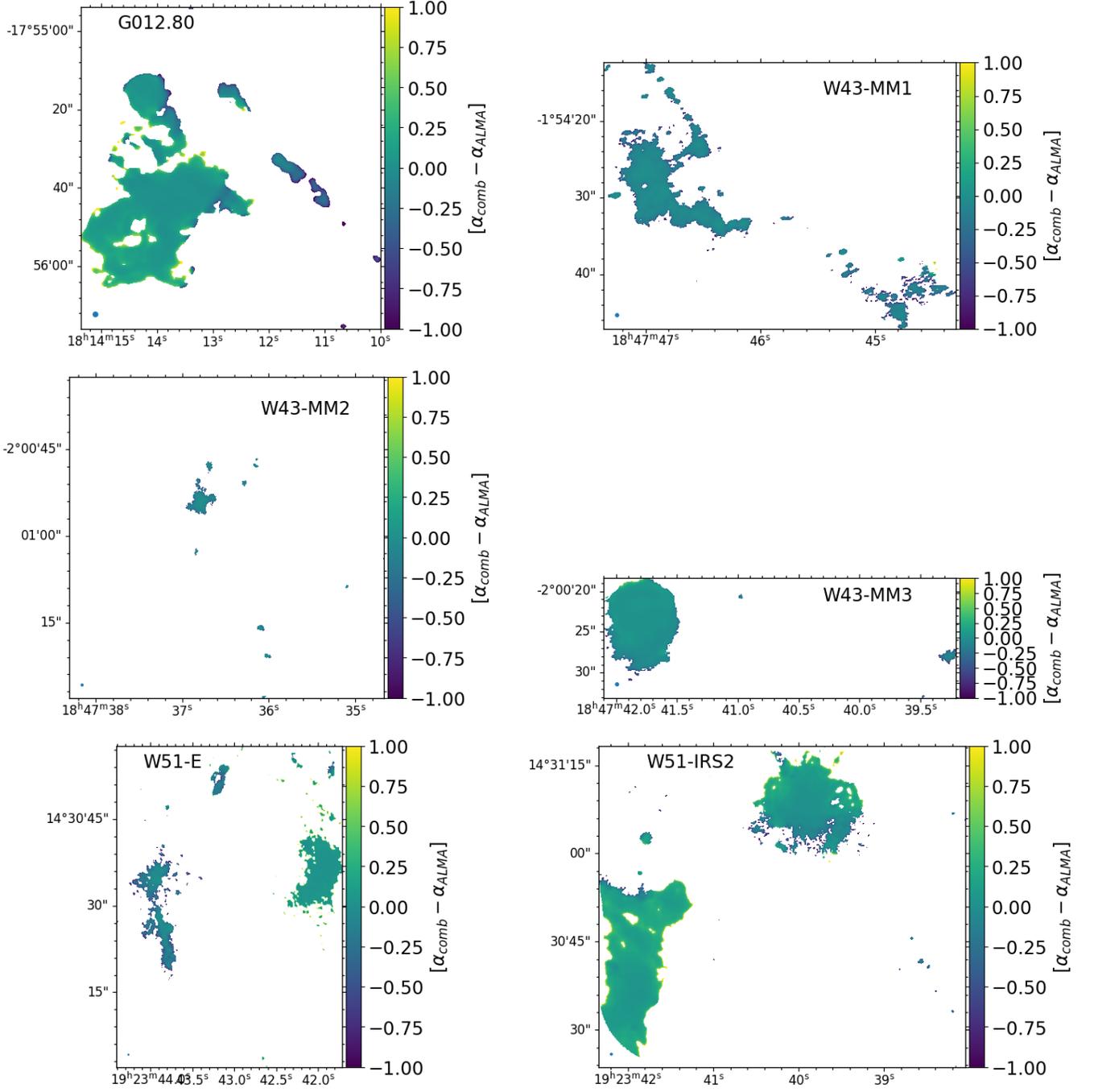

**Figure 13.** For each field, the panels show the difference $\alpha_{\mathrm{comb}} - \alpha_{\mathrm{ALMA}}$ between the spectral index measurements using the combined images and the *bsensnobright* interferometric images from the release of Ginsburg et al. (2022).

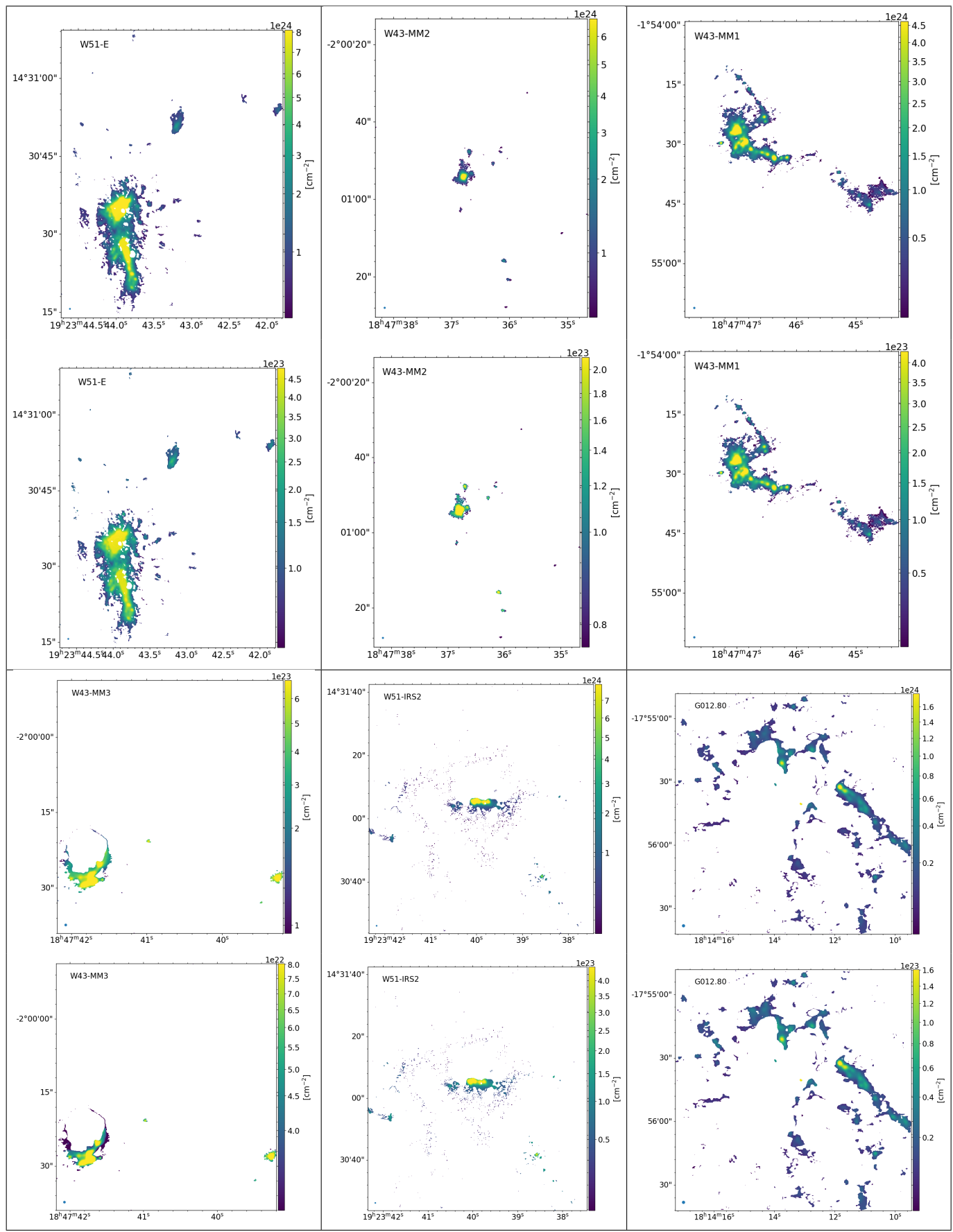

**Figure 14.** Molecular gas column density ($N_{\rm gas}$) maps and molecuar gas comlumn density error maps for each of the six regions regions with spectral index measurement. First and third row show the $N_{\rm gas}$ maps. Second and forth row show the $N_{\rm gas}$ error maps.

# D. DUST AND FREE-FREE DERIVED QUANTITIES

Table 9 shows the observed and derived quantities for the dust-dominated structures, based on the dendrogram identification at 1 mm (6 fields: G012.80, W43-MM1, W43-MM2, W43-MM3, W51-E, W51-IRS2). Table 10 shows the observed and derived quantities for the ionized structures, based on the dendrogram identification at 3 mm (4 fields: G012.80, W43-MM3, W51-E, W51-IRS2).

**Table 9**. Observed and derived quantities for dust-dominated pixels within dendrogram structures

| id | $S_{1mm}$ [mJy] | $D$ [pc] | $\Omega$ ['' $^2$] | $T_{dust}$ [K] | $\tau$ | $M_{gas}$ [$M_\odot$] | $<N_{gas}>$ [$\times 10^{23}$cm$^{-2}$] |
|---|---|---|---|---|---|---|---|
| | | | | G012.80 (distance = 2.4 kpc) | | | |
| T000 | $150 \pm 2$ | 0.07 | 27.97 | $25 \pm 2$ | $0.0070 \pm 0.0009$ | $13 \pm 2$ | $1.4 \pm 0.2$ |
| T001 | $127 \pm 1$ | 0.06 | 17.65 | $26 \pm 2$ | $0.009 \pm 0.001$ | $10 \pm 1$ | $1.8 \pm 0.2$ |
| T013 | $116 \pm 1$ | 0.06 | 18.61 | $27 \pm 2$ | $0.0073 \pm 0.0009$ | $9 \pm 1$ | $1.5 \pm 0.2$ |
| T008 | $356 \pm 3$ | 0.10 | 60.03 | $25 \pm 2$ | $0.0078 \pm 0.0010$ | $30 \pm 4$ | $1.7 \pm 0.2$ |
| T008-L011 | $58 \pm 1$ | 0.05 | 11.96 | $25 \pm 2$ | $0.0065 \pm 0.0008$ | $5.0 \pm 0.6$ | $2.9 \pm 0.3$ |
| T008-L018 | $38 \pm 1$ | 0.03 | 6.42 | $25 \pm 2$ | $0.0078 \pm 0.0010$ | $3.2 \pm 0.4$ | $3.3 \pm 0.4$ |
| T019 | $921 \pm 5$ | 0.18 | 182.95 | $29 \pm 2$ | $0.0054 \pm 0.0007$ | $64 \pm 8$ | $1.2 \pm 0.1$ |
| T019-L020 | $49 \pm 1$ | 0.05 | 12.39 | $27 \pm 2$ | $0.0048 \pm 0.0006$ | $3.8 \pm 0.5$ | $2.2 \pm 0.3$ |
| T019-L029 | $124 \pm 1$ | 0.03 | 3.81 | $29 \pm 2$ | $0.037 \pm 0.005$ | $9 \pm 1$ | $5.0 \pm 0.6$ |
| T019-L032 | $58.8 \pm 0.8$ | 0.03 | 3.79 | $29 \pm 2$ | $0.017 \pm 0.002$ | $4.2 \pm 0.5$ | $4.8 \pm 0.5$ |
| T019-L033 | $1006 \pm 2$ | 0.08 | 36.87 | $31 \pm 3$ | $0.028 \pm 0.003$ | $67 \pm 7$ | $7.0 \pm 0.7$ |
| T036 | $20.6 \pm 0.8$ | 0.03 | 3.81 | $30 \pm 3$ | $0.0057 \pm 0.0007$ | $1.4 \pm 0.2$ | $1.2 \pm 0.1$ |
| T037 | $35.2 \pm 0.9$ | 0.03 | 5.72 | $27 \pm 2$ | $0.0073 \pm 0.0009$ | $2.7 \pm 0.3$ | $1.6 \pm 0.2$ |
| T038 | $212 \pm 2$ | 0.07 | 30.32 | $28 \pm 2$ | $0.0081 \pm 0.0010$ | $16 \pm 2$ | $1.7 \pm 0.2$ |
| T002 | $5416 \pm 7$ | 0.24 | 336.33 | $32 \pm 3$ | $0.016 \pm 0.002$ | $350 \pm 40$ | $1.0 \pm 0.1$ |
| T002-L003 | $20 \pm 1$ | 0.05 | 16.59 | $29 \pm 2$ | $0.0013 \pm 0.0002$ | $1.4 \pm 0.2$ | $1.2 \pm 0.1$ |
| T002-L007 | $14.1 \pm 0.9$ | 0.03 | 4.62 | $30 \pm 3$ | $0.0031 \pm 0.0004$ | $0.9 \pm 0.1$ | $1.6 \pm 0.2$ |
| T002-L030 | $97.0 \pm 0.4$ | 0.01 | 1.04 | $37 \pm 3$ | $0.080 \pm 0.010$ | $5.4 \pm 0.5$ | $17 \pm 1$ |
| T002-L023 | $10.9 \pm 0.6$ | 0.02 | 2.11 | $35 \pm 3$ | $0.0045 \pm 0.0006$ | $0.62 \pm 0.08$ | $1.8 \pm 0.2$ |
| T002-L035 | $20.1 \pm 0.4$ | 0.01 | 1.04 | $35 \pm 3$ | $0.017 \pm 0.002$ | $1.2 \pm 0.1$ | $4.5 \pm 0.5$ |
| T002-L043 | $549 \pm 1$ | 0.06 | 18.18 | $35 \pm 3$ | $0.027 \pm 0.003$ | $32 \pm 3$ | $6.3 \pm 0.6$ |
| T002-L047 | $190 \pm 1$ | 0.04 | 10.08 | $34 \pm 3$ | $0.017 \pm 0.002$ | $11 \pm 1$ | $3.9 \pm 0.4$ |
| T002-L048 | $135 \pm 1$ | 0.04 | 9.09 | $32 \pm 3$ | $0.014 \pm 0.002$ | $8 \pm 1$ | $3.3 \pm 0.4$ |
| T002-L052 | $75 \pm 1$ | 0.04 | 10.47 | $31 \pm 3$ | $0.0071 \pm 0.0009$ | $4.8 \pm 0.6$ | $2.2 \pm 0.3$ |
| T002-L051 | $105 \pm 1$ | 0.05 | 16.69 | $31 \pm 3$ | $0.0062 \pm 0.0007$ | $6.7 \pm 0.8$ | $2.2 \pm 0.2$ |
| T002-L046 | $27.8 \pm 1.0$ | 0.03 | 5.95 | $29 \pm 2$ | $0.0050 \pm 0.0006$ | $1.9 \pm 0.2$ | $2.1 \pm 0.2$ |
| T002-L049 | $134 \pm 1$ | 0.05 | 13.67 | $29 \pm 2$ | $0.011 \pm 0.001$ | $9 \pm 1$ | $3.1 \pm 0.3$ |
| T002-L053 | $35 \pm 1$ | 0.04 | 9.80 | $27 \pm 2$ | $0.0043 \pm 0.0006$ | $2.7 \pm 0.3$ | $1.9 \pm 0.2$ |
| T050 | $47 \pm 1$ | 0.03 | 4.59 | $26 \pm 2$ | $0.013 \pm 0.002$ | $3.8 \pm 0.5$ | $1.5 \pm 0.2$ |
| T054 | $62 \pm 1$ | 0.03 | 4.84 | $25 \pm 2$ | $0.017 \pm 0.002$ | $5.3 \pm 0.6$ | $2.0 \pm 0.2$ |
| T055 | $40.6 \pm 0.9$ | 0.03 | 5.58 | $24 \pm 2$ | $0.010 \pm 0.001$ | $3.6 \pm 0.4$ | $2.1 \pm 0.3$ |
| | | | | W43-MM1 (distance = 5.5 kpc) | | | |
| T000 | $7.2 \pm 0.4$ | 0.03 | 1.30 | $22 \pm 2$ | $0.008 \pm 0.001$ | $3.6 \pm 0.5$ | $1.4 \pm 0.2$ |
| T001 | $10.4 \pm 0.4$ | 0.04 | 1.54 | $22 \pm 2$ | $0.010 \pm 0.001$ | $5.4 \pm 0.7$ | $1.9 \pm 0.2$ |
| T002 | $357 \pm 2$ | 0.27 | 80.91 | $21 \pm 2$ | $0.0071 \pm 0.0009$ | $190 \pm 20$ | $1.5 \pm 0.2$ |
| T002-L015 | $27.3 \pm 0.6$ | 0.06 | 4.05 | $21 \pm 2$ | $0.011 \pm 0.001$ | $15 \pm 2$ | $3.7 \pm 0.5$ |
| T002-L018 | $7.7 \pm 0.3$ | 0.03 | 0.83 | $22 \pm 2$ | $0.014 \pm 0.002$ | $4.1 \pm 0.5$ | $4.5 \pm 0.5$ |
| | | | | | | | Continued on next page |

**Table 9** – continued from previous page

| id | $S_{1mm}$ [mJy] | $D$ [pc] | $\Omega$ [$''^2$] | $T_{dust}$ [K] | $\tau$ | $M_{gas}$ [$M_\odot$] | $<N_{gas}>$ [$\times 10^{23}\mathrm{cm}^{-2}$] |
|---|---|---|---|---|---|---|---|
| T002-L019 | $13.3 \pm 0.3$ | 0.03 | 0.86 | $22 \pm 2$ | $0.024 \pm 0.003$ | $7.0 \pm 0.9$ | $4.8 \pm 0.6$ |
| T002-L012 | $3.2 \pm 0.3$ | 0.03 | 0.89 | $21 \pm 2$ | $0.0059 \pm 0.0010$ | $1.8 \pm 0.3$ | $2.4 \pm 0.3$ |
| T002-L009 | $5.0 \pm 0.2$ | 0.02 | 0.53 | $21 \pm 2$ | $0.015 \pm 0.002$ | $2.8 \pm 0.4$ | $4.8 \pm 0.6$ |
| T002-L011 | $69.4 \pm 0.6$ | 0.06 | 4.43 | $21 \pm 2$ | $0.025 \pm 0.003$ | $38 \pm 5$ | $6.9 \pm 0.8$ |
| T002-L014 | $17.2 \pm 0.3$ | 0.02 | 0.67 | $21 \pm 2$ | $0.042 \pm 0.006$ | $10 \pm 1$ | $6.6 \pm 0.8$ |
| T002-L017 | $16.0 \pm 0.4$ | 0.04 | 2.06 | $21 \pm 2$ | $0.013 \pm 0.002$ | $9 \pm 1$ | $4.0 \pm 0.5$ |
| T002-L020 | $45.3 \pm 0.7$ | 0.07 | 6.13 | $21 \pm 2$ | $0.012 \pm 0.002$ | $25 \pm 3$ | $4.1 \pm 0.5$ |
| T002-L022 | $3.0 \pm 0.3$ | 0.03 | 0.77 | $21 \pm 2$ | $0.0063 \pm 0.0010$ | $1.6 \pm 0.3$ | $2.9 \pm 0.4$ |
| T002-L029 | $2.2 \pm 0.3$ | 0.02 | 0.36 | $21 \pm 2$ | $0.010 \pm 0.002$ | $1.2 \pm 0.2$ | $2.2 \pm 0.3$ |
| T002-L030 | $16.7 \pm 0.4$ | 0.03 | 1.30 | $21 \pm 2$ | $0.021 \pm 0.003$ | $9 \pm 1$ | $5.1 \pm 0.6$ |
| T021 | $3.0 \pm 0.2$ | 0.02 | 0.62 | $20 \pm 2$ | $0.008 \pm 0.001$ | $1.7 \pm 0.3$ | $1.7 \pm 0.2$ |
| T037 | $6.8 \pm 0.3$ | 0.03 | 1.08 | $19 \pm 1$ | $0.011 \pm 0.002$ | $4.2 \pm 0.6$ | $1.8 \pm 0.2$ |
| T045 | $40.5 \pm 0.9$ | 0.09 | 9.55 | $21 \pm 2$ | $0.0067 \pm 0.0009$ | $22 \pm 3$ | $1.4 \pm 0.2$ |
| T045-L046 | $3.6 \pm 0.4$ | 0.04 | 1.58 | $21 \pm 2$ | $0.0036 \pm 0.0006$ | $2.0 \pm 0.3$ | $2.1 \pm 0.3$ |
| T045-L051 | $10.1 \pm 0.5$ | 0.06 | 3.45 | $21 \pm 2$ | $0.0046 \pm 0.0006$ | $5.4 \pm 0.7$ | $2.4 \pm 0.3$ |
| T062 | $4.7 \pm 0.3$ | 0.03 | 0.86 | $20 \pm 2$ | $0.009 \pm 0.001$ | $2.7 \pm 0.4$ | $2.0 \pm 0.3$ |
| T065 | $7.6 \pm 0.3$ | 0.03 | 0.97 | $23 \pm 2$ | $0.011 \pm 0.002$ | $3.7 \pm 0.5$ | $1.7 \pm 0.2$ |
| T023 | $975 \pm 4$ | 0.46 | 235.64 | $21 \pm 2$ | $0.0065 \pm 0.0008$ | $520 \pm 70$ | $1.4 \pm 0.2$ |
| T023-L070 | $4.7 \pm 0.4$ | 0.04 | 1.84 | $24 \pm 2$ | $0.0035 \pm 0.0005$ | $2.2 \pm 0.3$ | $1.9 \pm 0.2$ |
| T023-L027 | $3.2 \pm 0.3$ | 0.03 | 0.95 | $20 \pm 2$ | $0.0057 \pm 0.0009$ | $1.8 \pm 0.3$ | $2.7 \pm 0.3$ |
| T023-L032 | $5.4 \pm 0.3$ | 0.03 | 1.03 | $21 \pm 2$ | $0.009 \pm 0.001$ | $3.0 \pm 0.4$ | $2.8 \pm 0.3$ |
| T023-L042 | $5.1 \pm 0.2$ | 0.03 | 0.75 | $19 \pm 1$ | $0.012 \pm 0.002$ | $3.1 \pm 0.4$ | $4.2 \pm 0.5$ |
| T023-L072 | $11.6 \pm 0.3$ | 0.02 | 0.57 | $21 \pm 2$ | $0.032 \pm 0.004$ | $6.2 \pm 0.8$ | $4.8 \pm 0.6$ |
| T023-L049 | $38.8 \pm 0.2$ | 0.03 | 0.73 | $21 \pm 2$ | $0.09 \pm 0.01$ | $22 \pm 2$ | $20 \pm 2$ |
| T023-L050 | $73.8 \pm 0.3$ | 0.02 | 0.33 | $21 \pm 2$ | $0.42 \pm 0.07$ | $48 \pm 3$ | $27 \pm 2$ |
| T023-L047 | $303.0 \pm 0.4$ | 0.04 | 2.05 | $21 \pm 2$ | $0.26 \pm 0.04$ | $180 \pm 20$ | $56 \pm 4$ |
| T023-L053 | $242.7 \pm 0.5$ | 0.05 | 3.30 | $22 \pm 2$ | $0.12 \pm 0.02$ | $140 \pm 10$ | $26 \pm 2$ |
| T023-L057 | $136.6 \pm 0.3$ | 0.03 | 0.90 | $22 \pm 2$ | $0.27 \pm 0.04$ | $82 \pm 7$ | $39 \pm 3$ |
| T023-L058 | $1113.3 \pm 0.8$ | 0.09 | 8.50 | $22 \pm 2$ | $0.23 \pm 0.03$ | $650 \pm 60$ | $49 \pm 3$ |
| T023-L061 | $2273.5 \pm 0.9$ | 0.09 | 9.42 | $22 \pm 2$ | $0.45 \pm 0.07$ | $1450 \pm 90$ | $74 \pm 4$ |
| T023-L064 | $14.1 \pm 0.3$ | 0.03 | 0.97 | $23 \pm 2$ | $0.021 \pm 0.003$ | $6.9 \pm 0.8$ | $5.7 \pm 0.7$ |
| T023-L068 | $34.2 \pm 0.2$ | 0.03 | 0.62 | $24 \pm 2$ | $0.08 \pm 0.01$ | $17 \pm 2$ | $18 \pm 1$ |
| T023-L069 | $96.1 \pm 0.3$ | 0.03 | 0.89 | $24 \pm 2$ | $0.16 \pm 0.02$ | $49 \pm 4$ | $24 \pm 2$ |
| T023-L073 | $17.8 \pm 0.3$ | 0.03 | 0.83 | $23 \pm 2$ | $0.031 \pm 0.004$ | $9 \pm 1$ | $7.8 \pm 0.9$ |
| T023-L071 | $8.9 \pm 0.3$ | 0.03 | 0.83 | $21 \pm 2$ | $0.017 \pm 0.002$ | $4.8 \pm 0.6$ | $5.1 \pm 0.6$ |
| T023-L075 | $8.6 \pm 0.2$ | 0.02 | 0.64 | $21 \pm 2$ | $0.019 \pm 0.003$ | $4.3 \pm 0.5$ | $5.5 \pm 0.6$ |
| T023-L074 | $6.5 \pm 0.2$ | 0.03 | 0.73 | $22 \pm 2$ | $0.013 \pm 0.002$ | $3.3 \pm 0.4$ | $4.1 \pm 0.5$ |
| T023-L060 | $5.1 \pm 0.3$ | 0.03 | 0.80 | $23 \pm 2$ | $0.009 \pm 0.001$ | $2.5 \pm 0.3$ | $3.3 \pm 0.4$ |
| T023-L059 | $3.8 \pm 0.2$ | 0.03 | 0.71 | $23 \pm 2$ | $0.008 \pm 0.001$ | $1.9 \pm 0.3$ | $3.0 \pm 0.4$ |
| T023-L054 | $67.1 \pm 0.6$ | 0.06 | 4.55 | $20 \pm 2$ | $0.026 \pm 0.003$ | $40 \pm 5$ | $7.1 \pm 0.8$ |
| T023-L079 | $39.0 \pm 0.5$ | 0.05 | 2.88 | $22 \pm 2$ | $0.021 \pm 0.003$ | $21 \pm 3$ | $5.7 \pm 0.7$ |
| T023-L083 | $10.8 \pm 0.3$ | 0.03 | 0.87 | $21 \pm 2$ | $0.020 \pm 0.003$ | $6.0 \pm 0.7$ | $5.8 \pm 0.7$ |
| T023-L084 | $23.9 \pm 0.4$ | 0.04 | 1.82 | $20 \pm 2$ | $0.022 \pm 0.003$ | $14 \pm 2$ | $5.6 \pm 0.7$ |
| T023-L088 | $22.5 \pm 0.5$ | 0.06 | 3.41 | $20 \pm 2$ | $0.011 \pm 0.002$ | $13 \pm 2$ | $3.9 \pm 0.5$ |
| T023-L089 | $11.3 \pm 0.5$ | 0.05 | 2.86 | $19 \pm 1$ | $0.0071 \pm 0.0010$ | $6.9 \pm 0.9$ | $3.1 \pm 0.4$ |
| T076 | $6.1 \pm 0.3$ | 0.02 | 0.42 | $21 \pm 2$ | $0.024 \pm 0.003$ | $3.4 \pm 0.5$ | $2.0 \pm 0.3$ |

**Table 9** – continued from previous page

| id | $S_{1mm}$ | $D$ | $\Omega$ | $T_{dust}$ | $\tau$ | $M_{gas}$ | $<N_{gas}>$ |
|---|---|---|---|---|---|---|---|
| | [mJy] | [pc] | [$''^2$] | [K] | | [$M_\odot$] | [$\times 10^{23} cm^{-2}$] |
| T080 | $6.0 \pm 0.2$ | 0.03 | 0.76 | $21 \pm 2$ | $0.013 \pm 0.002$ | $3.3 \pm 0.4$ | $2.7 \pm 0.3$ |
| T085 | $11.1 \pm 0.5$ | 0.05 | 2.73 | $20 \pm 2$ | $0.0071 \pm 0.0010$ | $6.6 \pm 0.9$ | $1.4 \pm 0.2$ |
| T085-L086 | $5.8 \pm 0.3$ | 0.03 | 0.97 | $20 \pm 2$ | $0.010 \pm 0.001$ | $3.4 \pm 0.5$ | $3.0 \pm 0.4$ |
| T085-L087 | $1.2 \pm 0.3$ | 0.02 | 0.32 | $19 \pm 1$ | $0.007 \pm 0.002$ | $0.7 \pm 0.2$ | $1.9 \pm 0.2$ |
| T090 | $4.9 \pm 0.3$ | 0.01 | 0.11 | $18 \pm 1$ | $0.09 \pm 0.01$ | $3.3 \pm 0.4$ | $2.0 \pm 0.3$ |
| | | | W43-MM2 | | | (distance = 5.5 kpc) | |
| T000 | $26.0 \pm 0.3$ | 0.08 | 7.65 | $20 \pm 2$ | $0.0056 \pm 0.0007$ | $15 \pm 2$ | $1.2 \pm 0.2$ |
| T000-L001 | $5.6 \pm 0.2$ | 0.05 | 2.26 | $21 \pm 2$ | $0.0041 \pm 0.0005$ | $3.1 \pm 0.4$ | $2.0 \pm 0.3$ |
| T000-L002 | $7.2 \pm 0.2$ | 0.04 | 1.56 | $20 \pm 2$ | $0.008 \pm 0.001$ | $4.1 \pm 0.5$ | $2.7 \pm 0.3$ |
| T003 | $3.9 \pm 0.1$ | 0.02 | 0.53 | $20 \pm 2$ | $0.013 \pm 0.002$ | $2.3 \pm 0.3$ | $1.4 \pm 0.2$ |
| T004 | $45.7 \pm 0.3$ | 0.07 | 4.86 | $19 \pm 1$ | $0.017 \pm 0.002$ | $28 \pm 4$ | $3.7 \pm 0.5$ |
| T005 | $8.4 \pm 0.2$ | 0.04 | 2.07 | $20 \pm 2$ | $0.0072 \pm 0.0009$ | $5.0 \pm 0.7$ | $1.4 \pm 0.2$ |
| T006 | $124.6 \pm 0.7$ | 0.17 | 32.70 | $18 \pm 1$ | $0.0073 \pm 0.0010$ | $80 \pm 10$ | $1.6 \pm 0.2$ |
| T006-L007 | $75.3 \pm 0.4$ | 0.09 | 9.06 | $18 \pm 1$ | $0.016 \pm 0.002$ | $49 \pm 6$ | $5.0 \pm 0.6$ |
| T006-L009 | $4.5 \pm 0.2$ | 0.04 | 1.44 | $18 \pm 1$ | $0.0061 \pm 0.0008$ | $3.0 \pm 0.4$ | $2.7 \pm 0.3$ |
| T008 | $41.4 \pm 0.4$ | 0.10 | 11.29 | $18 \pm 1$ | $0.0071 \pm 0.0009$ | $27 \pm 4$ | $1.5 \pm 0.2$ |
| T011 | $13.5 \pm 0.2$ | 0.04 | 1.81 | $18 \pm 1$ | $0.014 \pm 0.002$ | $9 \pm 1$ | $2.7 \pm 0.3$ |
| T010 | $33.4 \pm 0.4$ | 0.09 | 9.23 | $20 \pm 2$ | $0.0061 \pm 0.0008$ | $19 \pm 2$ | $1.3 \pm 0.2$ |
| T014 | $14.1 \pm 0.2$ | 0.05 | 3.10 | $19 \pm 1$ | $0.008 \pm 0.001$ | $9 \pm 1$ | $1.7 \pm 0.2$ |
| T012 | $734 \pm 1$ | 0.45 | 227.34 | $21 \pm 2$ | $0.0051 \pm 0.0006$ | $390 \pm 50$ | $1.1 \pm 0.1$ |
| T012-L024 | $1.4 \pm 0.1$ | 0.03 | 1.14 | $23 \pm 2$ | $0.0018 \pm 0.0003$ | $0.7 \pm 0.1$ | $1.3 \pm 0.2$ |
| T012-L016 | $37.6 \pm 0.3$ | 0.07 | 6.09 | $21 \pm 2$ | $0.010 \pm 0.001$ | $21 \pm 3$ | $3.3 \pm 0.4$ |
| T012-L020 | $2.9 \pm 0.1$ | 0.03 | 0.77 | $21 \pm 2$ | $0.0059 \pm 0.0008$ | $1.6 \pm 0.2$ | $2.4 \pm 0.3$ |
| T012-L028 | $38.5 \pm 0.2$ | 0.05 | 3.12 | $20 \pm 2$ | $0.022 \pm 0.003$ | $23 \pm 3$ | $5.7 \pm 0.7$ |
| T012-L031 | $7.5 \pm 0.1$ | 0.03 | 1.00 | $21 \pm 2$ | $0.012 \pm 0.002$ | $4.2 \pm 0.5$ | $3.1 \pm 0.4$ |
| T012-L023 | $905.2 \pm 0.3$ | 0.08 | 7.18 | $24 \pm 2$ | $0.19 \pm 0.03$ | $470 \pm 40$ | $44 \pm 3$ |
| T012-L025 | $26.19 \pm 0.09$ | 0.02 | 0.58 | $23 \pm 2$ | $0.067 \pm 0.009$ | $13 \pm 1$ | $15 \pm 1$ |
| T012-L029 | $36.5 \pm 0.1$ | 0.03 | 0.80 | $23 \pm 2$ | $0.067 \pm 0.009$ | $18 \pm 2$ | $10 \pm 1$ |
| T012-L033 | $97.8 \pm 0.3$ | 0.07 | 5.47 | $22 \pm 2$ | $0.028 \pm 0.004$ | $53 \pm 6$ | $7.1 \pm 0.8$ |
| T012-L030 | $23.7 \pm 0.2$ | 0.06 | 3.51 | $19 \pm 1$ | $0.013 \pm 0.002$ | $15 \pm 2$ | $4.0 \pm 0.5$ |
| T012-L034 | $13.9 \pm 0.1$ | 0.03 | 0.73 | $19 \pm 1$ | $0.035 \pm 0.005$ | $9 \pm 1$ | $6.0 \pm 0.7$ |
| T012-L035 | $3.84 \pm 0.09$ | 0.02 | 0.52 | $19 \pm 1$ | $0.014 \pm 0.002$ | $2.4 \pm 0.3$ | $4.2 \pm 0.5$ |
| T036 | $9.9 \pm 0.2$ | 0.04 | 2.20 | $19 \pm 1$ | $0.008 \pm 0.001$ | $6.3 \pm 0.8$ | $1.7 \pm 0.2$ |
| T037 | $6.6 \pm 0.1$ | 0.03 | 1.19 | $19 \pm 1$ | $0.010 \pm 0.001$ | $4.2 \pm 0.6$ | $1.8 \pm 0.2$ |
| T038 | $12.0 \pm 0.2$ | 0.04 | 2.15 | $21 \pm 2$ | $0.009 \pm 0.001$ | $6.5 \pm 0.8$ | $1.8 \pm 0.2$ |
| | | W43-MM3 (distance = 5.5 kpc) | | | | | |
| T000 | $2.32 \pm 0.05$ | 0.02 | 0.59 | $18 \pm 1$ | $0.008 \pm 0.001$ | $1.5 \pm 0.2$ | $0.9 \pm 0.1$ |
| T011 | $1.66 \pm 0.04$ | 0.03 | 0.73 | $21 \pm 2$ | $0.0037 \pm 0.0005$ | $0.9 \pm 0.1$ | $0.78 \pm 0.10$ |
| T028 | $12.7 \pm 0.1$ | 0.07 | 5.05 | $23 \pm 2$ | $0.0037 \pm 0.0005$ | $6.3 \pm 0.8$ | $0.77 \pm 0.10$ |
| T001 | $870.7 \pm 1.0$ | 0.63 | 436.46 | $23 \pm 2$ | $0.0028 \pm 0.0004$ | $420 \pm 50$ | $0.57 \pm 0.07$ |
| T001-L008 | $6.4 \pm 0.1$ | 0.07 | 5.24 | $21 \pm 2$ | $0.0020 \pm 0.0003$ | $3.5 \pm 0.5$ | $1.0 \pm 0.1$ |
| T001-L010 | $8.33 \pm 0.09$ | 0.06 | 3.91 | $20 \pm 2$ | $0.0036 \pm 0.0005$ | $4.7 \pm 0.6$ | $1.4 \pm 0.2$ |
| T001-L018 | $19.5 \pm 0.2$ | 0.10 | 10.31 | $21 \pm 2$ | $0.0031 \pm 0.0004$ | $11 \pm 1$ | $1.3 \pm 0.2$ |
| T001-L025 | $1.71 \pm 0.04$ | 0.02 | 0.64 | $22 \pm 2$ | $0.0041 \pm 0.0005$ | $0.9 \pm 0.1$ | $1.5 \pm 0.2$ |
| T001-L024 | $18.48 \pm 0.10$ | 0.06 | 4.35 | $22 \pm 2$ | $0.0066 \pm 0.0008$ | $10 \pm 1$ | $2.0 \pm 0.2$ |
| T001-L030 | $169.7 \pm 0.2$ | 0.14 | 21.36 | $23 \pm 2$ | $0.012 \pm 0.001$ | $90 \pm 10$ | $3.1 \pm 0.4$ |

**Table 9 – continued from previous page**

| id | $S_{1mm}$ [mJy] | $D$ [pc] | $\Omega$ [$''^2$] | $T_{dust}$ [K] | $\tau$ | $M_{gas}$ [$M_\odot$] | $<N_{gas}>$ [$\times 10^{23} cm^{-2}$] |
|---|---|---|---|---|---|---|---|
| T001-L027 | $1.71 \pm 0.04$ | 0.03 | 0.82 | $22 \pm 2$ | $0.0031 \pm 0.0004$ | $0.9 \pm 0.1$ | $1.2 \pm 0.2$ |
| T001-L022 | $1.03 \pm 0.04$ | 0.03 | 0.70 | $22 \pm 2$ | $0.0023 \pm 0.0003$ | $0.54 \pm 0.07$ | $1.1 \pm 0.1$ |
| T001-L005 | $14.1 \pm 0.1$ | 0.09 | 8.18 | $21 \pm 2$ | $0.0028 \pm 0.0004$ | $7 \pm 1$ | $1.2 \pm 0.2$ |
| T001-L006 | $8.7 \pm 0.1$ | 0.07 | 5.38 | $21 \pm 2$ | $0.0026 \pm 0.0003$ | $4.8 \pm 0.6$ | $1.2 \pm 0.1$ |
| T001-L029 | $3.77 \pm 0.08$ | 0.05 | 2.52 | $24 \pm 2$ | $0.0020 \pm 0.0003$ | $1.7 \pm 0.2$ | $0.9 \pm 0.1$ |
| T001-L023 | $6.57 \pm 0.06$ | 0.04 | 1.49 | $24 \pm 2$ | $0.0060 \pm 0.0008$ | $3.1 \pm 0.4$ | $1.6 \pm 0.2$ |
| T001-L038 | $16.93 \pm 0.05$ | 0.02 | 0.47 | $27 \pm 2$ | $0.044 \pm 0.005$ | $7.1 \pm 0.8$ | $5.0 \pm 0.6$ |
| T001-L035 | $19.13 \pm 0.03$ | 0.02 | 0.52 | $27 \pm 2$ | $0.046 \pm 0.006$ | $8.0 \pm 0.9$ | $10 \pm 1$ |
| T001-L037 | $100.97 \pm 0.05$ | 0.03 | 1.01 | $27 \pm 2$ | $0.13 \pm 0.02$ | $43 \pm 4$ | $20 \pm 1$ |
| T001-L039 | $39.89 \pm 0.04$ | 0.03 | 0.69 | $28 \pm 2$ | $0.068 \pm 0.008$ | $16 \pm 2$ | $14 \pm 1$ |
| T001-L045 | $35.48 \pm 0.05$ | 0.02 | 0.62 | $29 \pm 2$ | $0.065 \pm 0.008$ | $14 \pm 1$ | $8.6 \pm 0.9$ |
| T001-L049 | $2.76 \pm 0.02$ | 0.01 | 0.15 | $29 \pm 2$ | $0.021 \pm 0.003$ | $1.0 \pm 0.1$ | $4.8 \pm 0.5$ |
| T001-L043 | $5.80 \pm 0.04$ | 0.02 | 0.58 | $28 \pm 2$ | $0.011 \pm 0.001$ | $2.3 \pm 0.3$ | $2.9 \pm 0.3$ |
| T001-L050 | $37.3 \pm 0.2$ | 0.10 | 11.20 | $25 \pm 2$ | $0.0042 \pm 0.0005$ | $16 \pm 2$ | $1.4 \pm 0.2$ |
| T001-L040 | $6.87 \pm 0.10$ | 0.06 | 4.29 | $25 \pm 2$ | $0.0020 \pm 0.0003$ | $3.0 \pm 0.4$ | $0.9 \pm 0.1$ |
| T001-L041 | $8.3 \pm 0.2$ | 0.12 | 15.43 | $22 \pm 2$ | $0.0008 \pm 0.0001$ | $4.3 \pm 0.5$ | $0.75 \pm 0.09$ |
| T051 | $4.85 \pm 0.07$ | 0.04 | 1.98 | $21 \pm 2$ | $0.0039 \pm 0.0005$ | $2.6 \pm 0.3$ | $0.76 \pm 0.10$ |
| T053 | $3.25 \pm 0.06$ | 0.03 | 1.32 | $20 \pm 2$ | $0.0043 \pm 0.0006$ | $1.9 \pm 0.3$ | $0.8 \pm 0.1$ |
| T054 | $13.1 \pm 0.1$ | 0.07 | 5.18 | $22 \pm 2$ | $0.0039 \pm 0.0005$ | $7.0 \pm 0.9$ | $0.8 \pm 0.1$ |
| T055 | $7.53 \pm 0.08$ | 0.05 | 2.87 | $20 \pm 2$ | $0.0044 \pm 0.0006$ | $4.3 \pm 0.6$ | $0.9 \pm 0.1$ |
| T056 | $10.0 \pm 0.1$ | 0.06 | 4.62 | $19 \pm 1$ | $0.0039 \pm 0.0005$ | $6.2 \pm 0.8$ | $0.8 \pm 0.1$ |
| W51-E (distance = 5.4 kpc) | | | | | | | |
| T000 | $10.8 \pm 0.6$ | 0.01 | 0.25 | $23 \pm 2$ | $0.064 \pm 0.009$ | $5.2 \pm 0.6$ | $7.4 \pm 0.9$ |
| T005 | $65 \pm 1$ | 0.04 | 2.06 | $24 \pm 2$ | $0.044 \pm 0.006$ | $30 \pm 3$ | $9 \pm 1$ |
| T001 | $3950 \pm 10$ | 0.38 | 162.27 | $30 \pm 3$ | $0.026 \pm 0.003$ | $1300 \pm 100$ | $6.0 \pm 0.7$ |
| T001-L027 | $4.0 \pm 0.8$ | 0.03 | 0.79 | $29 \pm 2$ | $0.006 \pm 0.001$ | $1.4 \pm 0.3$ | $6.7 \pm 0.7$ |
| T001-L023 | $6.6 \pm 0.4$ | 0.02 | 0.27 | $29 \pm 2$ | $0.028 \pm 0.004$ | $2.4 \pm 0.3$ | $11 \pm 1$ |
| T001-L044 | $16.3 \pm 0.6$ | 0.02 | 0.26 | $30 \pm 3$ | $0.069 \pm 0.009$ | $5.9 \pm 0.6$ | $13 \pm 1$ |
| T001-L034 | $16.9 \pm 0.5$ | 0.01 | 0.13 | $30 \pm 3$ | $0.15 \pm 0.02$ | $6.1 \pm 0.6$ | $15 \pm 1$ |
| T001-L043 | $11.7 \pm 0.4$ | 0.01 | 0.24 | $32 \pm 3$ | $0.049 \pm 0.006$ | $3.8 \pm 0.4$ | $15 \pm 1$ |
| T001-L047 | $22.4 \pm 0.6$ | 0.01 | 0.25 | $31 \pm 3$ | $0.09 \pm 0.01$ | $7.7 \pm 0.7$ | $16 \pm 1$ |
| T001-L017 | $67.7 \pm 0.6$ | 0.01 | 0.21 | $28 \pm 2$ | $0.46 \pm 0.07$ | $31 \pm 2$ | $48 \pm 3$ |
| T001-L022 | $74.5 \pm 0.5$ | 0.01 | 0.12 | $29 \pm 2$ | $1.1 \pm 0.2$ | $43 \pm 1$ | $54 \pm 3$ |
| T001-L024 | $147.5 \pm 0.6$ | 0.02 | 0.41 | $30 \pm 3$ | $0.47 \pm 0.07$ | $64 \pm 3$ | $73 \pm 4$ |
| T001-L026 | $87.7 \pm 0.5$ | 0.02 | 0.30 | $30 \pm 3$ | $0.36 \pm 0.05$ | $35 \pm 2$ | $80 \pm 4$ |
| T001-L028 | $4404 \pm 2$ | 0.07 | 6.13 | $32 \pm 3$ | $1.2 \pm 0.3$ | $2450 \pm 60$ | $77 \pm 4$ |
| T001-L031 | $145.7 \pm 0.8$ | 0.03 | 0.84 | $31 \pm 3$ | $0.19 \pm 0.03$ | $53 \pm 4$ | $43 \pm 3$ |
| T001-L032 | $148.5 \pm 0.8$ | 0.03 | 0.89 | $31 \pm 3$ | $0.18 \pm 0.02$ | $54 \pm 4$ | $41 \pm 3$ |
| T001-L037 | $29.9 \pm 0.4$ | 0.01 | 0.24 | $32 \pm 3$ | $0.13 \pm 0.02$ | $10.1 \pm 0.9$ | $33 \pm 2$ |
| T001-L046 | $53.6 \pm 0.7$ | 0.02 | 0.50 | $32 \pm 3$ | $0.11 \pm 0.01$ | $18 \pm 2$ | $24 \pm 2$ |
| T001-L051 | $68.8 \pm 0.8$ | 0.03 | 0.93 | $30 \pm 3$ | $0.079 \pm 0.010$ | $24 \pm 2$ | $21 \pm 1$ |
| T001-L045 | $22.3 \pm 0.6$ | 0.01 | 0.23 | $31 \pm 3$ | $0.10 \pm 0.01$ | $7.7 \pm 0.7$ | $16 \pm 1$ |
| T001-L060 | $18.1 \pm 0.5$ | 0.02 | 0.36 | $27 \pm 2$ | $0.060 \pm 0.008$ | $7.1 \pm 0.8$ | $19 \pm 1$ |
| T001-L057 | $8.4 \pm 0.5$ | 0.02 | 0.32 | $28 \pm 2$ | $0.029 \pm 0.004$ | $3.1 \pm 0.4$ | $12 \pm 1$ |
| T001-L035 | $11.5 \pm 0.9$ | 0.03 | 0.93 | $31 \pm 3$ | $0.013 \pm 0.002$ | $3.9 \pm 0.5$ | $7.8 \pm 0.8$ |
| T001-L054 | $17.6 \pm 1.0$ | 0.03 | 1.29 | $28 \pm 2$ | $0.015 \pm 0.002$ | $6.5 \pm 0.8$ | $9.0 \pm 1.0$ |

**Table 9 – continued from previous page**

| id | $S_{1mm}$ | $D$ | $\Omega$ | $T_{dust}$ | $\tau$ | $M_{gas}$ | $<N_{gas}>$ |
|---|---|---|---|---|---|---|---|
| | [mJy] | [pc] | [''²] | [K] | | [M$_\odot$] | [×10²³cm⁻²] |
| T033 | 53 ± 1 | 0.04 | 1.49 | 29 ± 2 | 0.040 ± 0.005 | 19 ± 2 | 8.2 ± 0.9 |
| T048 | 16.5 ± 0.6 | 0.02 | 0.30 | 30 ± 3 | 0.060 ± 0.008 | 5.8 ± 0.6 | 7.8 ± 0.8 |
| T040 | 182 ± 2 | 0.07 | 6.18 | 27 ± 2 | 0.035 ± 0.004 | 71 ± 8 | 7.5 ± 0.8 |
| T050 | 51 ± 1 | 0.04 | 1.43 | 28 ± 2 | 0.042 ± 0.005 | 19 ± 2 | 8.6 ± 0.9 |
| T052 | 53.3 ± 1.0 | 0.03 | 1.23 | 29 ± 2 | 0.048 ± 0.006 | 19 ± 2 | 9 ± 1 |
| T055 | 27.0 ± 0.8 | 0.03 | 0.82 | 28 ± 2 | 0.039 ± 0.005 | 10 ± 1 | 7.5 ± 0.8 |
| T059 | 7.7 ± 0.4 | 0.02 | 0.27 | 27 ± 2 | 0.035 ± 0.005 | 3.0 ± 0.4 | 7.4 ± 0.8 |
| T056 | 130 ± 1 | 0.06 | 4.45 | 24 ± 2 | 0.041 ± 0.005 | 60 ± 7 | 8.7 ± 1.0 |
| T029 | 2778 ± 3 | 0.11 | 15.00 | 27 ± 2 | 0.25 ± 0.03 | 1230 ± 90 | 5.9 ± 0.7 |
| T029-L061 | 2 ± 1 | 0.04 | 2.24 | 27 ± 2 | 0.0015 ± 0.0007 | 1.1 ± 0.5 | 6.2 ± 0.7 |
| T062 | 10.9 ± 0.5 | 0.02 | 0.35 | 25 ± 2 | 0.042 ± 0.006 | 4.8 ± 0.6 | 9 ± 1 |
| T063 | 323 ± 2 | 0.09 | 9.75 | 26 ± 2 | 0.043 ± 0.005 | 140 ± 20 | 8.6 ± 1.0 |
| T063-L064 | 114 ± 1 | 0.06 | 4.23 | 26 ± 2 | 0.035 ± 0.004 | 48 ± 5 | 15 ± 1 |
| T063-L066 | 5.5 ± 0.6 | 0.02 | 0.26 | 26 ± 2 | 0.027 ± 0.004 | 2.3 ± 0.4 | 11 ± 1 |
| T065 | 132 ± 1 | 0.06 | 3.57 | 27 ± 2 | 0.044 ± 0.006 | 52 ± 6 | 9 ± 1 |
| T067 | 77 ± 1 | 0.04 | 1.96 | 27 ± 2 | 0.048 ± 0.006 | 31 ± 3 | 6.4 ± 0.7 |
| T067-L069 | 4.6 ± 0.6 | 0.02 | 0.44 | 27 ± 2 | 0.013 ± 0.002 | 1.8 ± 0.3 | 8.8 ± 1.0 |
| T070 | 1.2 ± 0.2 | 0.01 | 0.05 | 27 ± 2 | 0.032 ± 0.006 | 0.49 ± 0.09 | 6.8 ± 0.8 |
| | | | | W51-IRS2 (distance = 5.4 kpc) | | | |
| T001 | 1907 ± 2 | 0.48 | 266.94 | 28 ± 2 | 0.0080 ± 0.0010 | 700 ± 80 | 1.7 ± 0.2 |
| T001-L032 | 5.0 ± 0.1 | 0.03 | 0.93 | 28 ± 2 | 0.0061 ± 0.0008 | 1.9 ± 0.2 | 2.7 ± 0.3 |
| T001-L006 | 8.7 ± 0.1 | 0.03 | 0.92 | 28 ± 2 | 0.011 ± 0.001 | 3.4 ± 0.4 | 3.6 ± 0.4 |
| T001-L013 | 61.5 ± 0.4 | 0.09 | 9.12 | 28 ± 2 | 0.0077 ± 0.0009 | 23 ± 3 | 3.4 ± 0.4 |
| T001-L021 | 27.6 ± 0.2 | 0.04 | 1.49 | 28 ± 2 | 0.021 ± 0.003 | 10 ± 1 | 5.8 ± 0.7 |
| T001-L026 | 8.7 ± 0.1 | 0.02 | 0.64 | 28 ± 2 | 0.016 ± 0.002 | 3.3 ± 0.4 | 5.1 ± 0.6 |
| T001-L029 | 81.1 ± 0.2 | 0.06 | 3.73 | 28 ± 2 | 0.025 ± 0.003 | 30 ± 3 | 6.9 ± 0.8 |
| T001-L033 | 121.3 ± 0.3 | 0.07 | 5.29 | 29 ± 2 | 0.025 ± 0.003 | 44 ± 5 | 7.1 ± 0.8 |
| T001-L036 | 12.8 ± 0.1 | 0.02 | 0.39 | 29 ± 2 | 0.036 ± 0.004 | 4.6 ± 0.5 | 5.0 ± 0.6 |
| T001-L041 | 11.22 ± 0.09 | 0.02 | 0.45 | 29 ± 2 | 0.027 ± 0.003 | 4.0 ± 0.4 | 7.5 ± 0.8 |
| T001-L045 | 25.51 ± 0.09 | 0.02 | 0.50 | 29 ± 2 | 0.057 ± 0.007 | 9 ± 1 | 13 ± 1 |
| T001-L046 | 210.8 ± 0.2 | 0.04 | 2.11 | 29 ± 2 | 0.11 ± 0.01 | 79 ± 7 | 24 ± 2 |
| T001-L043 | 21.5 ± 0.1 | 0.03 | 0.79 | 29 ± 2 | 0.030 ± 0.004 | 7.8 ± 0.9 | 6.1 ± 0.7 |
| T001-L047 | 42.8 ± 0.2 | 0.04 | 1.99 | 29 ± 2 | 0.023 ± 0.003 | 15 ± 2 | 6.3 ± 0.7 |
| T001-L052 | 61.2 ± 0.2 | 0.05 | 2.69 | 29 ± 2 | 0.025 ± 0.003 | 22 ± 2 | 6.8 ± 0.7 |
| T001-L053 | 10.1 ± 0.1 | 0.03 | 0.80 | 29 ± 2 | 0.014 ± 0.002 | 3.6 ± 0.4 | 4.6 ± 0.5 |
| T001-L008 | 33.5 ± 0.3 | 0.06 | 3.93 | 29 ± 2 | 0.010 ± 0.001 | 13 ± 2 | 3.9 ± 0.4 |
| T001-L056 | 19.4 ± 0.3 | 0.06 | 4.32 | 29 ± 2 | 0.0048 ± 0.0006 | 6.9 ± 0.8 | 2.7 ± 0.3 |
| T001-L031 | 10.1 ± 0.3 | 0.07 | 5.97 | 27 ± 2 | 0.0020 ± 0.0002 | 3.9 ± 0.5 | 2.2 ± 0.3 |
| T062 | 34.6 ± 0.2 | 0.06 | 3.60 | 30 ± 3 | 0.010 ± 0.001 | 12 ± 1 | 2.1 ± 0.2 |
| T063 | 3370 ± 1 | 0.45 | 234.92 | 34 ± 3 | 0.013 ± 0.002 | 1000 ± 100 | 1.4 ± 0.2 |
| T063-L071 | 14.4 ± 0.4 | 0.08 | 7.88 | 31 ± 3 | 0.0018 ± 0.0002 | 4.7 ± 0.6 | 1.9 ± 0.2 |
| T063-L086 | 23.5 ± 0.3 | 0.06 | 4.81 | 31 ± 3 | 0.0048 ± 0.0006 | 7.6 ± 0.9 | 2.5 ± 0.3 |
| T063-L118 | 66.8 ± 0.3 | 0.07 | 6.21 | 32 ± 3 | 0.011 ± 0.001 | 22 ± 2 | 3.7 ± 0.4 |
| T063-L097 | 19.7 ± 0.1 | 0.02 | 0.53 | 36 ± 3 | 0.032 ± 0.004 | 5.6 ± 0.6 | 5.0 ± 0.5 |
| T063-L091 | 12.85 ± 0.10 | 0.02 | 0.56 | 34 ± 3 | 0.021 ± 0.002 | 3.8 ± 0.4 | 5.8 ± 0.6 |
| T063-L098 | 13.1 ± 0.1 | 0.02 | 0.62 | 35 ± 3 | 0.019 ± 0.002 | 3.8 ± 0.4 | 5.3 ± 0.6 |
| | | | | | | | Continued on next page |

**Table 9** – continued from previous page

| id | $S_{1mm}$ | $D$ | $\Omega$ | $T_{dust}$ | $\tau$ | $M_{gas}$ | $<N_{gas}>$ |
|---|---|---|---|---|---|---|---|
| | [mJy] | [pc] | [$''^2$] | [K] | | [$M_\odot$] | [$\times 10^{23} cm^{-2}$] |
| T063-L105 | $94.9 \pm 0.1$ | 0.03 | 1.06 | $36 \pm 3$ | $0.077 \pm 0.009$ | $27 \pm 2$ | $14 \pm 1$ |
| T063-L112 | $454.5 \pm 0.1$ | 0.03 | 0.72 | $37 \pm 3$ | $0.7 \pm 0.1$ | $170 \pm 6$ | $117 \pm 4$ |
| T063-L116 | $257.10 \pm 0.09$ | 0.02 | 0.47 | $37 \pm 3$ | $0.58 \pm 0.09$ | $91 \pm 3$ | $115 \pm 4$ |
| T063-L110 | $4048.4 \pm 0.3$ | 0.06 | 4.23 | $37 \pm 3$ | $1.5 \pm 0.4$ | $2100 \pm 40$ | $114 \pm 4$ |
| T063-L103 | $88.4 \pm 0.2$ | 0.04 | 2.08 | $34 \pm 3$ | $0.038 \pm 0.005$ | $26 \pm 3$ | $9.0 \pm 0.9$ |
| T063-L104 | $25.1 \pm 0.1$ | 0.03 | 0.73 | $35 \pm 3$ | $0.030 \pm 0.004$ | $7.3 \pm 0.8$ | $7.8 \pm 0.8$ |
| T063-L108 | $53.4 \pm 0.1$ | 0.03 | 1.12 | $36 \pm 3$ | $0.042 \pm 0.005$ | $15 \pm 2$ | $8.5 \pm 0.9$ |
| T063-L115 | $42.9 \pm 0.2$ | 0.04 | 2.18 | $34 \pm 3$ | $0.018 \pm 0.002$ | $13 \pm 1$ | $4.9 \pm 0.5$ |
| T063-L137 | $1.10 \pm 0.03$ | 0.01 | 0.07 | $35 \pm 3$ | $0.014 \pm 0.002$ | $0.32 \pm 0.04$ | $4.4 \pm 0.5$ |
| T095 | $213.6 \pm 0.7$ | 0.16 | 30.80 | $28 \pm 2$ | $0.0077 \pm 0.0009$ | $78 \pm 9$ | $1.7 \pm 0.2$ |
| T095-L096 | $8.6 \pm 0.3$ | 0.07 | 6.22 | $28 \pm 2$ | $0.0016 \pm 0.0002$ | $3.2 \pm 0.4$ | $2.0 \pm 0.2$ |
| T095-L117 | $13.2 \pm 0.2$ | 0.03 | 1.15 | $29 \pm 2$ | $0.013 \pm 0.002$ | $4.8 \pm 0.6$ | $3.9 \pm 0.4$ |
| T095-L119 | $74.5 \pm 0.3$ | 0.06 | 3.97 | $29 \pm 2$ | $0.020 \pm 0.002$ | $27 \pm 3$ | $5.9 \pm 0.6$ |
| T106 | $49.1 \pm 0.3$ | 0.07 | 5.74 | $27 \pm 2$ | $0.010 \pm 0.001$ | $19 \pm 2$ | $2.1 \pm 0.3$ |
| T126 | $54.4 \pm 0.4$ | 0.08 | 7.62 | $30 \pm 3$ | $0.0075 \pm 0.0009$ | $19 \pm 2$ | $1.6 \pm 0.2$ |
| T126-L127 | $10.4 \pm 0.2$ | 0.06 | 3.69 | $30 \pm 3$ | $0.0029 \pm 0.0004$ | $3.6 \pm 0.4$ | $2.2 \pm 0.3$ |
| T126-L134 | $3.0 \pm 0.1$ | 0.03 | 1.08 | $29 \pm 2$ | $0.0030 \pm 0.0004$ | $1.1 \pm 0.1$ | $2.1 \pm 0.2$ |
| T130 | $56.5 \pm 0.3$ | 0.07 | 5.85 | $28 \pm 2$ | $0.011 \pm 0.001$ | $21 \pm 2$ | $2.3 \pm 0.3$ |

**Table 10.** Observed and derived quantities for free-free dominated pixels within dendrogram structures

| id | $S_{3mm}$ [mJy] | $\Omega$ $['^2]$ | $D$ [pc] | EM [$\times 10^6$ pc cm$^{-6}$] | $n_e$ [$\times 10^3$ cm$^{-3}$] | $Q_0$ [$\times 10^{47}$ s$^{-1}$] |
|---|---|---|---|---|---|---|
| G012.80 (distance = 2.4 kpc) | | | | | | |
| T0002 | $940 \pm 6$ | 993.99 | 0.41 | $0.59 \pm 0.10$ | $1.2 \pm 0.1$ | $3.1 \pm 0.5$ |
| T0002-L0036 | $4.2 \pm 0.3$ | 2.66 | 0.02 | $1.0 \pm 0.2$ | $6.7 \pm 0.6$ | $0.014 \pm 0.003$ |
| T0002-L0042 | $7.9 \pm 0.5$ | 5.35 | 0.03 | $0.9 \pm 0.2$ | $5.5 \pm 0.5$ | $0.026 \pm 0.005$ |
| T0002-L0007 | $54 \pm 1$ | 30.20 | 0.07 | $1.1 \pm 0.2$ | $3.9 \pm 0.3$ | $0.18 \pm 0.03$ |
| T0002-L0020 | $12.1 \pm 0.5$ | 6.04 | 0.03 | $1.2 \pm 0.2$ | $6.2 \pm 0.5$ | $0.040 \pm 0.007$ |
| T0002-L0035 | $11.3 \pm 0.4$ | 3.90 | 0.03 | $1.8 \pm 0.3$ | $8.3 \pm 0.7$ | $0.037 \pm 0.006$ |
| T0002-L0050 | $1326 \pm 1$ | 65.60 | 0.11 | $12 \pm 2$ | $10.9 \pm 0.9$ | $4.4 \pm 0.7$ |
| T0002-L0047 | $30.1 \pm 0.5$ | 4.68 | 0.03 | $4.0 \pm 0.7$ | $11.8 \pm 1.0$ | $0.10 \pm 0.02$ |
| T0002-L0040 | $225.6 \pm 0.9$ | 20.49 | 0.06 | $6 \pm 1$ | $10.7 \pm 0.9$ | $0.7 \pm 0.1$ |
| T0002-L0046 | $39.8 \pm 0.5$ | 4.91 | 0.03 | $5.0 \pm 0.8$ | $13 \pm 1$ | $0.13 \pm 0.02$ |
| T0002-L0048 | $48.3 \pm 0.5$ | 6.10 | 0.03 | $4.9 \pm 0.8$ | $12 \pm 1$ | $0.16 \pm 0.03$ |
| T0002-L0024 | $220.9 \pm 0.8$ | 14.74 | 0.05 | $9 \pm 1$ | $13 \pm 1$ | $0.7 \pm 0.1$ |
| T0002-L0019 | $158.3 \pm 0.7$ | 10.32 | 0.04 | $9 \pm 1$ | $15 \pm 1$ | $0.52 \pm 0.09$ |
| T0002-L0023 | $413.1 \pm 0.9$ | 18.81 | 0.06 | $13 \pm 2$ | $15 \pm 1$ | $1.4 \pm 0.2$ |
| T0002-L0028 | $185.5 \pm 0.6$ | 7.72 | 0.04 | $14 \pm 2$ | $20 \pm 1$ | $0.6 \pm 0.1$ |
| T0002-L0030 | $697 \pm 1$ | 24.51 | 0.06 | $17 \pm 2$ | $16 \pm 1$ | $2.3 \pm 0.4$ |
| T0002-L0033 | $3198.8 \pm 0.9$ | 20.46 | 0.06 | $96 \pm 16$ | $40 \pm 3$ | $10 \pm 1$ |
| T0002-L0038 | $410.5 \pm 0.4$ | 4.31 | 0.03 | $59 \pm 9$ | $46 \pm 3$ | $1.3 \pm 0.2$ |
| W43-MM3 (distance = 5.5 kpc) | | | | | | |
| T0000 | $66.7 \pm 0.8$ | 31.12 | 0.17 | $1.3 \pm 0.2$ | $2.8 \pm 0.2$ | $1.2 \pm 0.2$ |
| T0000-L0009 | $82.4 \pm 0.2$ | 2.68 | 0.05 | $19 \pm 3$ | $19 \pm 1$ | $1.4 \pm 0.2$ |
| T0000-L0007 | $93.4 \pm 0.2$ | 2.62 | 0.05 | $22 \pm 3$ | $21 \pm 1$ | $1.6 \pm 0.3$ |
| T0000-L0008 | $20.2 \pm 0.1$ | 0.62 | 0.02 | $20 \pm 3$ | $29 \pm 2$ | $0.35 \pm 0.06$ |
| W51-E (distance = 5.4 kpc) | | | | | | |
| T0012 | $788 \pm 2$ | 120.33 | 0.32 | $4.1 \pm 0.7$ | $3.5 \pm 0.3$ | $13 \pm 2$ |
| T0012-L0138 | $5.7 \pm 0.1$ | 0.30 | 0.02 | $11 \pm 1$ | $26 \pm 2$ | $0.09 \pm 0.02$ |
| T0012-L0057 | $46.4 \pm 0.3$ | 1.84 | 0.04 | $15 \pm 2$ | $19 \pm 1$ | $0.8 \pm 0.1$ |
| T0012-L0128 | $65.6 \pm 0.3$ | 2.35 | 0.05 | $17 \pm 2$ | $19 \pm 1$ | $1.1 \pm 0.2$ |
| T0012-L0125 | $150.9 \pm 0.4$ | 4.27 | 0.06 | $21 \pm 3$ | $18 \pm 1$ | $2.5 \pm 0.4$ |
| T0012-L0111 | $6.22 \pm 0.10$ | 0.21 | 0.01 | $18 \pm 3$ | $36 \pm 3$ | $0.10 \pm 0.02$ |
| T0012-L0074 | $6.92 \pm 0.10$ | 0.22 | 0.01 | $19 \pm 3$ | $37 \pm 3$ | $0.12 \pm 0.02$ |
| T0012-L0109 | $197.7 \pm 0.5$ | 5.43 | 0.07 | $22 \pm 3$ | $18 \pm 1$ | $3.3 \pm 0.6$ |
| T0012-L0120 | $47.1 \pm 0.2$ | 1.31 | 0.03 | $22 \pm 3$ | $25 \pm 2$ | $0.8 \pm 0.1$ |
| T0012-L0105 | $26.7 \pm 0.2$ | 0.53 | 0.02 | $31 \pm 5$ | $38 \pm 3$ | $0.44 \pm 0.07$ |
| T0012-L0101 | $31.8 \pm 0.1$ | 0.49 | 0.02 | $40 \pm 6$ | $44 \pm 3$ | $0.53 \pm 0.09$ |
| T0012-L0099 | $26.0 \pm 0.1$ | 0.39 | 0.02 | $41 \pm 6$ | $47 \pm 3$ | $0.43 \pm 0.07$ |
| T0012-L0100 | $22.8 \pm 0.1$ | 0.28 | 0.02 | $50 \pm 8$ | $56 \pm 4$ | $0.38 \pm 0.06$ |
| T0012-L0086 | $29.8 \pm 0.1$ | 0.34 | 0.02 | $54 \pm 9$ | $56 \pm 4$ | $0.49 \pm 0.08$ |
| T0012-L0094 | $25.2 \pm 0.1$ | 0.23 | 0.01 | $67 \pm 11$ | $69 \pm 5$ | $0.42 \pm 0.07$ |
| T0012-L0096 | $160.3 \pm 0.2$ | 1.38 | 0.03 | $71 \pm 12$ | $45 \pm 3$ | $2.7 \pm 0.4$ |
| T0033 | $2.6 \pm 0.1$ | 0.42 | 0.02 | $3.9 \pm 0.7$ | $14 \pm 1$ | $0.044 \pm 0.008$ |
| T0033-L0089 | $465.9 \pm 0.1$ | 0.23 | 0.01 | $1276 \pm 215$ | $300 \pm 25$ | $7 \pm 1$ |
| T0033-L0059 | $20.91 \pm 0.09$ | 0.19 | 0.01 | $68 \pm 11$ | $72 \pm 6$ | $0.35 \pm 0.06$ |

**Table 10 – continued from previous page**

| id | $S_{3mm}$ | $\Omega$ | $D$ | EM | $n_e$ | $Q_0$ |
|---|---|---|---|---|---|---|
| | [mJy] | $[''^2]$ | [pc] | $[\times 10^6 \text{ pc cm}^{-6}]$ | $[\times 10^3 \text{ cm}^{-3}]$ | $[\times 10^{47} \text{ s}^{-1}]$ |
| T0139 | $16.80 \pm 0.09$ | 0.20 | 0.01 | $52 \pm 8$ | $62 \pm 5$ | $0.28 \pm 0.05$ |
| | | | | W51-IRS2 (distance = 5.4 kpc) | | |
| T0029 | $54 \pm 1$ | 7.29 | 0.08 | $4.7 \pm 0.8$ | $7.6 \pm 0.6$ | $0.9 \pm 0.2$ |
| T0035 | $1064 \pm 5$ | 166.46 | 0.38 | $4.0 \pm 0.7$ | $3.2 \pm 0.3$ | $17 \pm 2$ |
| T0035-L0038 | $13.7 \pm 0.4$ | 1.22 | 0.03 | $6 \pm 1$ | $14 \pm 1$ | $0.23 \pm 0.04$ |
| T0035-L0062 | $50.4 \pm 0.7$ | 2.81 | 0.05 | $11 \pm 1$ | $14 \pm 1$ | $0.8 \pm 0.1$ |
| T0035-L0058 | $55.6 \pm 0.7$ | 3.69 | 0.06 | $9 \pm 1$ | $12 \pm 1$ | $0.9 \pm 0.2$ |
| T0035-L0061 | $21.4 \pm 0.4$ | 1.20 | 0.03 | $11 \pm 1$ | $18 \pm 1$ | $0.36 \pm 0.06$ |
| T0035-L0063 | $66.7 \pm 0.5$ | 1.47 | 0.04 | $28 \pm 4$ | $28 \pm 2$ | $1.1 \pm 0.2$ |
| T0035-L0055 | $153.4 \pm 0.5$ | 1.91 | 0.04 | $49 \pm 8$ | $34 \pm 2$ | $2.6 \pm 0.4$ |
| T0035-L0056 | $30.8 \pm 0.2$ | 0.34 | 0.02 | $56 \pm 9$ | $57 \pm 4$ | $0.51 \pm 0.09$ |
| T0035-L0060 | $2104 \pm 1$ | 12.00 | 0.10 | $108 \pm 18$ | $32 \pm 2$ | $35 \pm 5$ |
| T0064 | $26.1 \pm 0.7$ | 3.54 | 0.06 | $4.6 \pm 0.8$ | $9.1 \pm 0.8$ | $0.43 \pm 0.07$ |

## E. DATA PRODUCTS

The first set of products provided with this paper are the combined maps of MGPS90 with ALMA-IMF Band 3 and BGPS with ALMA-IMF Band 6. The *bsensnobright* continuum images released by Ginsburg et al. (2022) are used. The combined images at 3 mm have the `_combination.MGPS90.B3.fits` file extension. For the 1 mm combination they have the `_combination.BGPS.B6.fits` extension.

The second set of products includes the structures identified by the dendrogram algorithm. The structures identified in the combined images at 3 mm have the `DDRG.MGPS90.B3.fits` extension. For the 1 mm structures, the extension is `DDRG.BGPS.B6.fits`.

The third set of products comprises the spectral index maps (extension `.ALPHA.fits`) and their corresponding error maps (`.ALPHA.ERR.fits`).

The previously described products can be found in the following Zenodo repository: https://zenodo.org/record/8110640.